\documentclass[11pt,a4paper]{article}
\pdfoutput=1
\usepackage{jheppub_modified_green}
\usepackage{amsmath,amssymb,amsfonts}
\usepackage{multirow}
\usepackage{geometry}
\usepackage{array}
\usepackage{graphicx}
\usepackage{calc}
\usepackage{color}
\usepackage{hyperref}
\usepackage[usenames, dvipsnames]{xcolor}
\usepackage[normalem]{ulem}
\usepackage{caption}
\newcommand{\ym}[1]{\ensuremath{A_{\text{YM}}^{#1}}}

\title{On-Shell Gauge Invariant Three-Point Amplitudes}
\author[a,1]{Zhengdi Sun,}
\note{First Authors}
\author[a,1]{Hui Xu,}
\author[a,2]{Yeuk-Kwan E. Cheung,}
\note{Corresponding Author}

\affiliation[a]{School of Physics, Nanjing University, Nanjing, 210093, China}

\emailAdd{zdsun@smail.nju.edu.cn}
\emailAdd{huixu@smail.nju.edu.cn}
\emailAdd{cheung@nju.edu.cn}

\abstract{{
Assuming locality, Lorentz invariance and parity conservation we obtain a set of differential equations governing the  
3-point  interactions of massless bosons, 
 which in turn determines the polynomial ring of these amplitudes. 
We derive  all possible 3-point interactions for tensor fields with polarisations that have total symmetry 
 and  mixed symmetry  under permutations of Lorentz indices. 
Constraints on the existence of gauge-invariant cubic vertices for totally symmetric
 fields are obtained in general spacetime dimensions 
  and are compared with existing results obtained 
in   the covariant  and  light-cone approaches. \\
Expressing our results in spinor helicity formalism we reproduce the perhaps 
 mysterious mismatch between the covariant approach and the light cone approach in 4 dimensions. 
Our analysis also shows  that there exists a mismatch, in the 3-point gauge invariant amplitudes corresponding to cubic self-interactions,  between a scalar field $\phi$ and an antisymmetric rank-2 tensor field $A_{\mu\nu}$.  
 Despite the well-known fact
that  in 4 dimensions  rank-2 anti-symmetric  fields are dual to  scalar fields 
 in free theories, such duality does not extend to 
  interacting theories.}
}

\begin{document}
\maketitle
\flushbottom

\section{Introduction}

The gauge principle has evolved   over the past century from an insight of Weyl to being held as a guiding principle for constructing quantum theories that describe the interactions of elementary particles.
The current paradigm of particle physics is summed up in the Standard Model~(SM) with $SU(3)\times SU(2)\times U(1)$ gauge bosons interacting with the observed multiplets of quarks and leptons.
While there are numerous attempts to extend this paradigm,
 the most noticeable being supersymmetry~\cite{WESS197439, WESS197452}, there are also efforts to extend the gauge interactions  to higher spin fields~\cite{Bengtsson:1983pd, 
Bengtsson:1986kh, Metsaev:2005ar, Bekaert:2006py, Benincasa:2007xk, Sagnotti:2010at, Fotopoulos:2010ay, Joung:2011ww, Bengtsson:2014qza, Conde:2016izb, Boels:2016xhc, Ponomarev:2016lrm, Sleight:2016xqq}.
The difficulty of constructing interacting theories involving only  finitely many spins were already noticed in some of these early works~\cite{Berends:1984rq}. 
 The most recent attempts to extend the gauge sector to
infinitely  higher spins can be found in, e.g.~\cite{Vasiliev:1995dn, Savvidy:2008zy}; 
see also works in this directions from the string perspective~\cite{Sagnotti:2010at, Fotopoulos:2010ay}.

{
Since the 1980s  there have been various approaches to determine the possible 
higher-spin interactions, the first attempts being by 
 Bengtsson et al~\cite{Bengtsson:1983pd,  Bengtsson:1986kh}
in which three-point higher-spin vertices, unique for a particular set of spins, were obtained in the light-cone formalism. 
A more general result was later  obtained  in~\cite{Metsaev:2005ar}  by  a similar method,
in which Metsaev  used commutators of the
{Poincar\'{e}} algebra to obtain the parity-even cubic vertices for massless
fields in four dimensions, the cubic vertices for massless totally-symmetric fields in five dimensions and the cubic vertices for massless fields in six dimensions. 
The   covariant approach  was initiated  in~\cite{Berends:1984rq}, as they constructed 
the unique self-interactions for  massless spin-1, 2, 3 fields and  interactions 
of two scalars with a spin-$s$ boson using the  Fronsdal fields.
This line of research was  recently completed by 
Manvelyan \textit{et al}.~\cite{Manvelyan:2010jr}. 
Sagnotti \textit{et al}.~\cite{Sagnotti:2010at} and 
Fotopoulos \textit{et al}.~\cite{Fotopoulos:2010ay} 
derived the same results from the string theory.}

On the 3-point amplitudes Benincasa and Cachazo~\cite{Benincasa:2007xk} 
exploited the technique of spinor helicity to
derive a general form of helicity amplitudes for three massless particles.
 But this construction does not manifest the gauge symmetry, since spinors transform
  trivially under the translations (the gauge transformations in this case) 
 in the little group.  Therefore given a spinor helicity amplitude in~\cite{Benincasa:2007xk}  
   it is not clear whether there exists a Lagrangian description of a gauge theory 
which could lead to such an amplitude or not. It is, nevertheless, possible to include gauge transformations 
 in the spinor helicity formalism (for example, see~\cite{Elvang:2013cua}). 
Benincasa and Cachazo  used BCFW to investigate a class of ``constructible'' theories 
and  found that  there are  no nontrivial amplitudes of different species of spin-2 particles or particles with spin larger than 2 among this class of theories,
while a theory with a single kind of spin-2 particles is unique. 
Lately Boels and Medina~\cite{Boels:2016xhc} have obtained the three-point amplitudes using the constraints of on-shell gauge invariance. Their results  are expressed in terms of polarization tensors with the gauge transformations manifest;
these have  been done  for polarization vectors and rank-2 polarization tensors, but not for general polarizations.

{
There emerges a series of  works combining these two approaches to study  the spinor helicity amplitudes~\cite{Conde:2016izb, Sleight:2016xqq, Ponomarev:2016lrm}, 
the most notable discovery being   a mismatch between the light-cone and covariant 
 approaches: there are cubic vertices existing in light-cone approach but are absent in the covariant approach~\cite{Conde:2016izb, Sleight:2016xqq, Ponomarev:2016lrm, Bengtsson:2014qza}. 
 As it turns out, the missing part is crucial for the existence of the higher-spin 
 theory in 4d Minkowski spacetime, as pointed out by~\cite{Conde:2016izb} and~\cite{Ponomarev:2016lrm}. 
 }

{
In this paper we study the 3-point gauge-invariant amplitudes which are expressed in terms of polarization tensors. 
Lorentz invariance and locality give strong constraints 
on the amplitudes in gauge theories so it is natural to use these constraints 
to select the possible theories. We focus on the parity conserved theories. 

Let us  first clarify what we mean by gauge invariance. We use the term gauge in the situation where the descriptions in a theory have redundancies. For example two polarization vectors $e_\mu(p)$ and $e_\mu(p) + p_\mu$ describe the same physical state of  a massless spin-1 boson. 
This is a redundancy in the description. 
We use the term gauge invariance to refer to the fact that all physical quantities, such as scattering amplitudes,  do not depend on how we describe a particular physical state. 
Although in the following discussion, we only consider the variation of the polarization tensors, our discussion is still valid for Non-Abelian Gauge theories: at the zeroth order (in coupling constant $g$)  the gauge transformation does not change the color of the external states (the term containing $f^{abc}$ also has  a factor $g$).  
Therefore we can simply drop the factor $f^{abc}$ in the following discussion on the 
gauge invariance of 3-point amplitudes.

With  these theoretical  assumptions  we  start with    the 3-point gauge-invariant amplitudes of totally symmetric fields whose polarizations  can be written as
$\epsilon^{\pm}(p)=\bigotimes_{i=1}^{s}e^{\pm}(p)$.
We  find four basic gauge-invariant amplitudes, from  which all  possible the 3-point gauge-invariant amplitudes of three totally symmetric tensor fields  can be constructed. 
We are able to give constraints  on the total number of derivatives that are allowed to appear in a 3-point gauge-invariant vertex. 
Although we present our work in 3+1 dimensions, 
our analysis   of the  totally  symmetric fields is  valid for
d+1 dimensions ($d\geq 3$): since  changing the dimensions changes neither the form of polarization tensors nor 
their gauge transformations for totally  symmetric tensor~\cite{Bekaert:2006py}. 
Whereas  the  number of  polarization directions  can change, the  polarization tensors in this case  
are  tensor products  of  polarization vectors in  the same direction. 
 One  special thing, nevertheless, arises in four dimensions:  many of the otherwise allowed  amplitudes may vanish due to a Schouten-like identity.}

The general results are summarized as follows,
\begin{equation}
    \begin{aligned}
        & A(\epsilon_1,\epsilon_2,\epsilon_3;N)\\
        = & (e_1\cdot p_2)^{\frac{s_1-s_2-s_3+N}{2}}(e_2\cdot p_1)^{\frac{s_2-s_1-s_3+N}{2}}(e_3\cdot p_1)^{\frac{s_3-s_1-s_2+N}{2}}A_{\text{YM}}(e_1,e_2,e_3)^{\frac{s_1+s_2+s_3-N}{2}}
    \end{aligned}
\end{equation}
where $N$ is the total number of derivatives in the cubic vertex and
\begin{equation}
        A_{\text{YM}}(e_1,e_2,e_3)\equiv (e_1\cdot p_2)(e_2\cdot e_3)-(e_2\cdot p_1)(e_1\cdot e_3)+(e_3\cdot p_1)(e_1\cdot e_2)
\end{equation}

In 4-dimension, however, due to a Schouten type identity, the non-trivial amplitudes are given by (assuming $s_1\leq s_2\leq s_3$ for convenience)
\begin{equation}
	(e_1\cdot p_2)^{s_1} (e_2\cdot p_1)^{s_2} (e_3\cdot p_1)^{s_3}～，
\end{equation}
and
\begin{equation}
	(e_2\cdot p_1)^{s_2 - s_1} (e_3\cdot p_1)^{s_3 - s_1} A_{YM}^{s_1}~.
\end{equation}

Next we turn our attention to the 3-point gauge-invariant amplitudes of fields 
with mixed symmetries. We show that, like in the case of totally symmetric fields, 
there are certain basic amplitudes which can be used to construct  all other amplitudes; 
we provide a detailed recipe to do so. 
It is known (for example, see~\cite{Bekaert:2006py}) that 
all massless mixed-symmetry fields are dual to totally-symmetric fields in the free theory in 4-dimension. 
Our result shows  that this duality does not extend to the interacting theory: 
there exists a mismatch between the 3-point amplitudes upon introduction  of interaction. 
Expressing our results in spinor helicity formalism  we  show that 
in  two representations of the same helicity if the 3-point amplitudes for a given 
{set} of  momenta do exist in  both theories, 
 these two  amplitudes must be the same (up to coupling constants). 
 This is because the 3-point amplitudes in spinor helicity formalism only depend on 
 their  helicities. 
Also in the spinor-helicity formalism, we can derive the conditions for non-vanishing amplitudes.  This constraint is, however, missing  in the light-cone approach.
Therefore  we can reproduce, and perhaps help to elucidate, the mismatch between the 
covariant approach and the light-cone approach.

The article is organized as follows:
In the next section, we briefly review  polarizations and  gauge transformations
to set  notations. In Section~\ref{sec:amplitudes_set} we discuss the gauge
invariant 3-point amplitudes of higher-spin fields with  total  symmetry
and mixed  symmetry in their Lorentz indices.
 In Section~\ref{sec:examples}  we present an example of how to construct a
 specific amplitude. And Section~\ref{sec:conclusion} is a 
 brief conclusion { and a short discussion. 
 A straight forward derivation for all possible gauge invariant 3-point amplitudes 
 in the case of totally-symmetric tensors is presented in Appendix. }

\section{Polarizations and Gauge Transformations}

In the construction of a Hamiltonian with field operators that leads to a Lorentz 
invariant S-matrix, a generic field operator $\psi_l$ is required to transform
according to a representation of the Lorentz group~\cite{weinberg1995quantum, 
 Weinberg:1964ew}:
\begin{equation}  \label{field_operator_condition}
    U(\Lambda,a)\psi_l(x)U^{-1}(\Lambda,a)
    =\sum_{l'}D_{ll'}(\Lambda^{-1})\psi_{l'}(\Lambda x+a)
\end{equation}
where $U(\Lambda,a)$ is the operator corresponding to the
Poincar\'{e} transformation $x'=\Lambda x+a$. In the case of a
 rank-$r$ tensor field 
$\phi^{\mu_1\cdots\mu_r}$ describing a massless spin $s$ particle, the condition~(\ref{field_operator_condition}) is fulfilled if the field operator is of the form
\begin{equation}
\phi^{\mu_1\cdots\mu_r}
=\frac{1}{(2\pi)^{3/2}}\sum_{\sigma=\pm s}
  \int \frac{d^3p}{\sqrt{2p^0}} 
  \left[
      \epsilon^{\mu_1\cdots\mu_r}
   (\mathbf{p},\sigma)a(\mathbf{p},\sigma)e^{ip\cdot x}
    +\epsilon^{\mu_1\cdots\mu_r}(\mathbf{p},\sigma)^*
    {a^c}^\dagger(\mathbf{p},\sigma)e^{-ip\cdot x}
  \right]
\end{equation}
with the polarization tensor satisfying~\cite{weinberg1995quantum}
\begin{align}
    & {D\left[R(\theta)\right]^{\mu'_1\cdots\mu'_r}}_{\mu_1\cdots\mu_r}\epsilon^{\mu_1\cdots\mu_r} (\mathbf{k},\sigma) =\epsilon^{\mu'_1\cdots\mu'_r} (\mathbf{k},\sigma)e^{i\sigma\theta}\quad (\sigma=\pm s)\label{tensorCondition1}\\
    & {D\left[S(\alpha,\beta)\right]^{\mu'_1\cdots\mu'_r}}_{\mu_1\cdots\mu_r}\epsilon^{\mu_1\cdots\mu_r} (\mathbf{k},\sigma) =\epsilon^{\mu'_1\cdots\mu'_r} (\mathbf{k},\sigma)\label{tensorCondition2}
\end{align}
where $k=(1,0,0,1)$ is a standard momentum { and}  $R(\theta)$, 
$S(\alpha,\beta)$ are the little group transformations:
\begin{align}
    {R(\theta)^\mu}_\nu &
   =\begin{bmatrix}
         1 & 0 & 0 & 0\\
       0 & \cos\theta & \sin\theta & 0 \\
      0 & -\sin\theta & \cos\theta & 0\\
      0 & 0 & 0 & 1\\
  \end{bmatrix}\\
 {S(\alpha,\beta)^\mu}_\nu &
  =\begin{bmatrix}
  1-\dfrac{\alpha^2+\beta^2}{2} & \alpha\,\,\,\,
   & \beta & -\dfrac{\alpha^2+\beta^2}{2}\\
    \alpha & 1\,\,\,\, & 0 & -\alpha\\
    \beta & 0\,\,\,\, & 1 & -\beta\\
   \dfrac{\alpha^2+\beta^2}{2} & \alpha\,\,\,\,
   & \beta & 1-\dfrac{\alpha^2+\beta^2}{2}
\end{bmatrix}
\end{align}
and $D\left[R(\theta)\right],D\left[S(\alpha,\beta)\right]$ are the tensor representation matrices. { Equation (\ref{tensorCondition2}) requires that the translations of the little group act trivially to exclude continuous internal quantum numbers (For example, see~\cite{Bekaert:2006py}).}  Usually the above equations cannot be simultaneously satisfied. In such cases, we require only Equation~(\ref{tensorCondition1}) to hold and 
that any physical quantities (for example, the scattering amplitudes) must be invariant under  translations of the $ISO(2)$ (which we loosely call the ``gauge transformation''):
\begin{equation}\label{gaugeTransformation}
    \epsilon^{\mu'_1\cdots\mu'_r} (\mathbf{k},\sigma)\rightarrow {D\left[S(\alpha,\beta)\right]^{\mu'_1\cdots\mu'_r}}_{\mu_1\cdots\mu_r}\epsilon^{\mu_1\cdots\mu_r} (\mathbf{k},\sigma)
\end{equation}
which, nevertheless, ensures  Lorentz invariance of the amplitudes.

 In order to 
 solve~(\ref{tensorCondition1}) and determine the gauge transformation~(\ref{gaugeTransformation}),
let $J_z$ be the generator of the transformations $R(\theta)$ and
$I+\tilde{S}(\alpha,\beta)$ be the infinitesimal version of $S(\alpha,\beta)$:
\begin{equation}
    {(J_z)^\mu}_\nu=
    \begin{bmatrix}
    0 & 0 & 0 & 0\\
    0&0&-i&0\\
    0&i&0&0\\
    0&0&0&0
    \end{bmatrix},
\quad
{\tilde{S}(\alpha,\beta)^\mu}_\nu
=  \begin{bmatrix}
         0 & \alpha & \beta & 0\\
        \alpha & 0 & 0 & -\alpha\\
         \beta & 0 & 0 & -\beta\\
       0 & \alpha & \beta & 0\\
    \end{bmatrix}
\end{equation}
{
Then the tensor condition~(\ref{tensorCondition1}) and  the gauge transformation~(\ref{gaugeTransformation}) become
\begin{equation}\label{tensorConditionModified}
    (J_z\otimes I\otimes\cdots\otimes I\, + I\otimes J_z\otimes\cdots\otimes I\, 
    +\cdots + I\otimes \cdots\otimes I\otimes J_z) \epsilon(\mathbf{k},\sigma) 
       =\sigma \epsilon(\mathbf{k},\sigma)~,
\end{equation}
\begin{equation}\label{tran}
    \delta \epsilon(\mathbf{k},\sigma)=\left[\tilde{S}(\alpha,\beta)\otimes I\otimes\cdots\otimes I+I\otimes \tilde{S}(\alpha,\beta)\otimes\cdots\otimes I+\cdots+I\otimes \cdots\otimes I\otimes \tilde{S}(\alpha,\beta)\right] \epsilon(\mathbf{k},\sigma)~. \,\,
\end{equation}
}
The eigenvectors of $J_z$ are, in turn,  
$e^\mu(\mathbf{k},1)=(0,1,i,0)$, 
$e^\mu(\mathbf{k},-1)=(0,1,-i,0)$, 
$k^\mu=(1,0,0,1)$, 
$\bar{k}^\mu=(1,0,0,-1)$ 
with  the following properties:
\begin{equation}\label{basisProperties}
\begin{aligned}
    & J_z e(\mathbf{k},1)=e(\mathbf{k},1),\, J_z e(\mathbf{k},-1)=-e(\mathbf{k},-1),\, J_z k=J_z \bar{k}=0\\
    & \tilde{S}(\alpha,\beta)e(\mathbf{k},1)=\xi k,\,\tilde{S}(\alpha,\beta) e(\mathbf{k},-1)=\zeta k,\, \tilde{S}(\alpha,\beta) k=0,\\
    & \tilde{S}(\alpha,\beta) \bar{k}=\zeta e(\mathbf{k},1) +\xi e(\mathbf{k},-1)
\end{aligned}
\end{equation}
where $\xi\equiv \alpha+i\beta,\zeta\equiv \alpha-i\beta$.

The set of tensor products of $e(\mathbf{k},1)$, 
$e(\mathbf{k},-1),k$ and $\bar{k}$ 
forms a basis of the linear space of rank-$r$ tensors.
By expressing $\epsilon(\mathbf{k},\sigma)$ as a linear combination of these tensor
 products, using~(\ref{tensorConditionModified}) and~(\ref{basisProperties}), 
 we can thus obtain the general form of the polarization tensors:
\begin{equation}\label{polarization_expression}
    \epsilon(\mathbf{k},\sigma)=\sum_\pi \lambda_\pi \bigotimes_{i=1}^r e(\mathbf{k},\pi(i))
\end{equation}
where the functions $\pi$: $\{1,2,\cdots,r\}\rightarrow \{1,-1\}$
satisfy $\sum_{i=1}^r \pi(i)=\sigma$. 
The infinitesimal gauge transformations
are given by~(\ref{tran}):
\begin{equation}\label{gaugeT}
\begin{aligned}
    \delta_{\xi,\zeta}\epsilon(\mathbf{k},\sigma)
  &=\big[\xi(\Delta^+\otimes I\otimes\cdots\otimes I+I\otimes\Delta^+\otimes I\otimes\cdots\otimes I+\cdots+I\otimes\cdots\otimes I\otimes\Delta^+ )+\\
   &\phantom{=[}\zeta(\Delta^-\otimes I\otimes\cdots\otimes I+
   I\otimes\Delta^-\otimes I\otimes\cdots\otimes I
   +\cdots+I\otimes\cdots\otimes I\otimes\Delta^- )\big]\epsilon(\mathbf{k},\sigma)\\
   &\equiv (\xi \delta^+ +\zeta\delta^-) \epsilon(\mathbf{k},\sigma)~,
\end{aligned}
\end{equation}
with the linear operators $\Delta^+,\Delta^-$ defined by
\begin{align*}
    &\Delta^\pm e(\mathbf{k},\pm 1)=k,\quad\Delta^\pm e(\mathbf{k},\mp 1)=0~.
\end{align*}

\section{Gauge Invariant 3-Point Amplitudes }
\label{sec:amplitudes_set}

In this section, we shall use on-shell gauge invariance to determine the three-point amplitudes of massless bosons with integral spins. Lorentz invariance, locality and parity conservation are assumed throughout this paper.
We shall also assume that the invariance of amplitudes under  transformations~(\ref{gaugeT}) holds for complex momenta. 
This  may be a general property of the amplitude: analytic continuation in the momenta does not break the on-shell gauge invariance. We do not have a proof and 
we take it as an assumption. 
With complex momenta, $p_1,\,  p_2,\, p_3$ are not forced to be  collinear,  even though momentum conservation ($p_1+p_2+p_3=0$)
and the massless on-shell condition ($p_i^2=0$)  
imply $p_i\cdot p_j =0$.
As a result, $e_i\cdot p_j \neq 0$ for $i\neq j$ in general.

{In subsection~\ref{sec:totally_symmetric1}, we write down the general amplitudes for totally symmetric fields constructed from Lorentz invariant pieces and then compute (in Appendix) explicitly the variation of the amplitudes under gauge transformations~(\ref{gaugeT}). 
Demanding  the variations be zero, one can determine the coefficients in the amplitudes. 
This method is straight forward, but not appropriate to apply it to the case with  mixed symmetry. 
In subsections~\ref{sec:polynomial_ring} and~\ref{sec:amplitudes_mixed_symmetry} we turn the  on-shell gauge invariance  conditions into a set of differential equations to determine the amplitudes, which applies conveniently to both the totally symmetric case and the case with mixed-symmetry.}

\subsection{Amplitudes of Totally Symmetric Polarizations}
	\label{sec:totally_symmetric1}
Consider three massless particles whose polarization tensors
$\epsilon_1(p_1)$, $\epsilon_2(p_2)$ and  $\epsilon_3(p_3)$,
 with $(p_1+p_2+p_3=0)$,  are given by,
\begin{equation}\label{polarizations}
  \epsilon^{\pm}_1(p_1)
    =\bigotimes_{i=1}^{s_1}e^{\pm}_1(p_1),\quad \epsilon^{\pm}_2(p_2)
    =\bigotimes_{i=1}^{s_2}e^{\pm}_2(p_2),\quad \epsilon^{\pm}_3(p_3)
    =\bigotimes_{i=1}^{s_3}e^{\pm}_3(p_3)
\end{equation}
where $s_i$ denote the spins of the particles and, $e_i(p_i)$,   
the polarization vectors (sometimes denoted by 
$e_i^\pm(p_i){{\equiv e_i(\boldsymbol{p}_i,\pm s_i)}}$  to emphasize  the positive or the negative helicities 
respectively, $\pm$ is omitted when there is no ambiguity.).

A complete set of  gauge invariant amplitudes is obtained with a  straightforward calculation, presented  
in the~\hyperref[appendix]{Appendix}. 
Each independent amplitude is labelled by, $N$, the number of derivatives:
\begin{equation}\label{amp_sss}
    \begin{aligned}
        & A(\epsilon_1,\epsilon_2,\epsilon_3;N)\\
        = & (e_1\cdot p_2)^{\frac{s_1-s_2-s_3+N}{2}}(e_2\cdot p_1)^{\frac{s_2-s_1-s_3+N}{2}}(e_3\cdot p_1)^{\frac{s_3-s_1-s_2+N}{2}}A_{\text{YM}}(e_1,e_2,e_3)^{\frac{s_1+s_2+s_3-N}{2}}
    \end{aligned}
\end{equation}
where
\begin{equation}
        A_{\text{YM}}(e_1,e_2,e_3)\equiv (e_1\cdot p_2)(e_2\cdot e_3)-(e_2\cdot p_1)(e_1\cdot e_3)+(e_3\cdot p_1)(e_1\cdot e_2)
\end{equation}
and $N$ satisfies
\begin{equation} \label{eq:N}
    s_1+s_2+s_3-2\min(s_1,s_2,s_3)\leqslant N\leqslant s_1+s_2+s_3~.
\end{equation}

{ Note that in the above  discussion we consider only the 4-dimensional case.  
We can, nevertheless, generalize our results to any D-dimensions~($D\geq 4$), because our derivation in 
the~\hyperref[appendix]{Appendix} only depends on the transversality of the polarization vectors, the absence  of 
self-contractions $e_i\cdot e_i$ and the form of the on-shell gauge transformations, $\delta e_i\propto p_i$.  
These  still hold in dimensions larger than four, see, e.g. section 5.3.1 of~\cite{Bekaert:2006py}.
The range of allowed momenta,  of  the generalized results,   
agrees with the  corresponding results 
 in the  light-cone approach~\cite{Metsaev:2005ar} 
 and the results obtained in the covariant 
 approach~\cite{Manvelyan:2010jr, Sagnotti:2010at}.}

{ In 4-dimensions, however, there are only 4 linearly independent vectors.
 As a result, a Schouten-like identity  makes  some of the amplitudes acquired in 
 the generic dimensions  vanish. To see this, consider the following 5-by-5 matrix,
\begin{equation}
	M_{ij} = \begin{pmatrix}
 	& e_1\cdot e_1 & e_1\cdot e_2 & e_1 \cdot e_3 & e_1\cdot p_1 & e_1\cdot p_2 \\
 	& e_2\cdot e_1 & e_2\cdot e_2 & e_2 \cdot e_3 & e_2\cdot p_1 & e_2\cdot p_2 \\
 	& e_3\cdot e_1 & e_3\cdot e_2 & e_3 \cdot e_3 & e_3\cdot p_1 & e_3\cdot p_2 \\
 	& p_1\cdot e_1 & p_1\cdot e_2 & p_1 \cdot e_3 & p_1\cdot p_1 & p_1\cdot p_2 \\
 	& p_2\cdot e_1 & p_2\cdot e_2 & p_2 \cdot e_3 & p_2\cdot p_1 & p_2\cdot p_2 \\
 \end{pmatrix}
 	= \begin{pmatrix}
 		&0 & e_1\cdot e_2 & e_1\cdot e_3 & 0 & e_1\cdot p_2 \\
 		&e_2\cdot e_1 & 0 & e_2\cdot e_3 & e_2\cdot p_1 & 0 \\
 		&e_3\cdot e_1 & e_3\cdot e_2 & 0 & e_3\cdot p_1 & e_3\cdot p_2\\
 		&0 & p_1\cdot e_2 & p_1\cdot e_3 & 0 & 0 \\
 		&p_2\cdot e_1 & 0 & p_2\cdot e_3 & 0 & 0 \\
 	\end{pmatrix}~.
\end{equation}
Since only  4 of these vectors are linearly independent, the determinant of this 5-by-5 matrix must vanish:
\begin{equation}
	\det{M_{ij}} = -2(e_1\cdot p_2)(e_2\cdot p_1)(e_3\cdot p_1)\bigg[(e_1\cdot p_2)(e_2\cdot e_3) + (e_3\cdot e_1)(e_2\cdot p_3) + (e_1\cdot e_2)(e_3\cdot p_1)\bigg] = 0~.
\end{equation}
This implies that $(e_1\cdot p_2)\,(e_2\cdot p_1)\,(e_3\cdot p_1)\, A_{YM} = 0$, and thus
 the only non-vanishing amplitudes in 4D are:
\begin{equation}
	(e_1\cdot p_2)^{s_1} (e_2\cdot p_1)^{s_2} (e_3\cdot p_1)^{s_3}~,
\end{equation}
and
\begin{equation}
	(e_2\cdot p_1)^{s_2 - s_1} (e_3\cdot p_1)^{s_3 - s_1} A_{YM}^{s_1}~,
\end{equation}
 taking  $s_1\leq s_2\leq s_3$.

Another way to see this is to express the amplitudes in the spinor helicity formalism. 
The non-vanishing amplitudes satisfying~(\ref{eq:N}) are given by (taking  $s_1\leq s_2\leq s_3$):
\begin{equation}
	\begin{aligned}
	A(\epsilon_1^+,\epsilon_2^+,\epsilon_3^+;N=s_1+s_2+s_3) & \propto
	[12]^{s_1+s_2-s_3}[23]^{s_3+s_2-s_1}[31]^{s_1+s_3-s_2}\\
	A(\epsilon_1^-,\epsilon_2^-,\epsilon_3^-;N=s_1+s_2+s_3) & \propto \langle12\rangle^{s_1+s_2-s_3}\langle23\rangle^{s_3+s_2-s_1}\langle31
	\rangle^{s_1+s_3-s_2}\\
	A(\epsilon_2^+,\epsilon_3^+,\epsilon_1^-;N=s_2+s_3-s_1) & 
	\propto [23]^{s_2+s_3+s_1}[31]^{s_3-s_1-s_2}[ki]^{s_2-s_1-s_3}\\
    A(\epsilon_2^-,\epsilon_3^-,\epsilon_1^+;N=s_2+s_3-s_1) & 
    \propto \langle 23\rangle^{s_2+s_3+s_1}\langle 31\rangle^{s_3-s_1-s_2}
    \langle 12\rangle^{s_2-s_1-s_3}~.
\end{aligned}
\end{equation}
In a nutshell there can be only  1 type of bracket appearing in the   3-point amplitudes 
in 4-D due to momentum conservation; and it is not hard to see that   these non-vanishing  amplitudes are the only ones  satisfying the constraint. 
}

\subsection{Polynomial Ring of Gauge Invariant Amplitudes}
    \label{sec:polynomial_ring}
    
\subsubsection{The Totally Symmetric Case: Yet Another Way }
 We propose  a different method to otain~(\ref{amp_sss}). 
This method can be applied to the analysis of gauge invariant amplitudes of tensor fields with polarizations of mixed symmetry. 
Let us first  define
\begin{equation}
 \left\{\def \arraystretch{1.5}
     \begin{array}{lll}
    X_1\equiv e_1\cdot p_2, & X_2\equiv e_2\cdot p_3,
       & X_3\equiv e_3\cdot p_1,\\
    Y_1\equiv e_2\cdot e_3, & Y_2\equiv e_3\cdot e_1,
        & Y_3\equiv e_1\cdot e_2.
     \end{array}
 \right.
\end{equation}
The gauge invariance of the amplitude
$A[\bigotimes e_1,\bigotimes e_2,\bigotimes e_3]$ 
requires that
\begin{equation}
  \left\{
    \begin{aligned}
       & \left(X_3\frac{\partial}{\partial Y_2}-X_2
          \frac{\partial}{\partial Y_3}\right)A=0\\
       & \left(X_1\frac{\partial}{\partial Y_3}-X_3
           \frac{\partial}{\partial Y_1}\right)A=0\\
      & \left(X_2\frac{\partial}{\partial Y_1}-X_1
          \frac{\partial}{\partial Y_2}\right)A=0~.
     \end{aligned}
  \right.
\end{equation}
Setting $\mathbf{X}\equiv(X_1,X_2,X_3)$ and  
$\mathbf{Y}\equiv (Y_1,Y_2,Y_3)$,
these equations can be combined into a single vectorial  equation:
\begin{equation}
    \mathbf{X}\times \nabla_{\mathbf{Y}}A=0~,
\end{equation}
 yielding $\nabla_{\mathbf{Y}}A\propto \mathbf{X}\Rightarrow A
     = f_X(\mathbf{X}\cdot \mathbf{Y})$,
 where  $f_X$ is a functions that depends on $(X_1,X_2,X_3)$.
For this solution to be a proper amplitude, it should be a polynomial in
$\mathbf{X}$ and $\mathbf{Y}$. 
This requirement can be fulfilled iff
$A\in \mathbb{C}[\mathbf{X}][\mathbf{X}\cdot \mathbf{Y}]=\mathbb{C}[\mathbf{X},\mathbf{X}\cdot \mathbf{Y}]$, with $K[\xi]$ denoting a polynomial ring over $K$.
Note that $\mathbf{X}\cdot \mathbf{Y}$ is nothing but the Yang-Mills amplitude.
We thus  conclude that the full set of gauge invariant amplitudes
consists of all polynomials in  ($X_1,X_2,X_3$)
and  $\mathbf{X}\cdot \mathbf{Y}=A_{\text{YM}}(e_1,e_2,e_3)$,
consistent with~(\ref{amp_sss}).

\subsubsection{The Generic Case}

Similarly if we allow the polarizations
$\epsilon_1,\epsilon_2,\epsilon_3$
to be general~(\ref{polarization_expression}) with mixed symmetry.
By  definitions,
\begin{equation}
    \left\{\def \arraystretch{1.5}
        \begin{array}{lll}
            X_1^\pm\equiv e_1^\pm\cdot p_2, & X_2^\pm\equiv e_2^\pm\cdot p_3,
                & X_3^\pm\equiv e_3^\pm\cdot p_1,\\
            Y_1^{\pm,\pm}\equiv e_2^\pm\cdot e_3^\pm, & Y_2^{\pm,\pm}\equiv
              e_3^\pm\cdot e_1^\pm, & Y_3^{\pm,\pm}\equiv e_1^\pm\cdot e_2^\pm
        \end{array}
    \right.
\end{equation}
the gauge invariance conditions become:
\begin{equation}
 \left\{
  \begin{aligned}
    & \left(X_3^+\frac{\partial}{\partial Y_2^{++}}+X_3^-\frac{\partial}{\partial Y_2^{-+}}-X_2^+\frac{\partial}{\partial Y_3^{++}}-X_2^-\frac{\partial}{\partial Y_3^{+-}}\right)A=0\\
   & \left(X_3^+\frac{\partial}{\partial Y_2^{+-}}+X_3^-\frac{\partial}{\partial Y_2^{--}}-X_2^+\frac{\partial}{\partial Y_3^{-+}}-X_2^-\frac{\partial}{\partial Y_3^{--}}\right)A=0\\
   & \left(X_1^+\frac{\partial}{\partial Y_3^{++}}+X_1^-\frac{\partial}{\partial Y_3^{-+}}-X_3^+\frac{\partial}{\partial Y_1^{++}}-X_3^-\frac{\partial}{\partial Y_1^{+-}}\right)A=0\\
   & \left(X_1^+\frac{\partial}{\partial Y_3^{+-}}+X_1^-\frac{\partial}{\partial Y_3^{--}}-X_3^+\frac{\partial}{\partial Y_1^{-+}}-X_3^-\frac{\partial}{\partial Y_1^{--}}\right)A=0\\
   & \left(X_2^+\frac{\partial}{\partial Y_1^{++}}+X_2^-\frac{\partial}{\partial Y_1^{-+}}-X_1^+\frac{\partial}{\partial Y_1^{-+}}-X_1^-\frac{\partial}{\partial Y_2^{+-}}\right)A=0\\
   & \left(X_2^+\frac{\partial}{\partial Y_1^{+-}}+X_2^-\frac{\partial}{\partial Y_1^{--}}-X_1^+\frac{\partial}{\partial Y_2^{-+}}-X_1^-\frac{\partial}{\partial Y_2^{--}}\right)A=0
   \end{aligned}~.
  \right.
\end{equation}
We  have assumed that the amplitudes contain  no self 
contractions $e_i^+\cdot e_i^-$, without  loss of  generality. 

Only 5 of the above equations are linearly independent and
because the number of $Y$'s is 12,  there are 7 independent solutions for $A$.
By  ``independent solutions"  we mean the functions of  $Y$'s  whose degrees of freedom lie  along the independent directions in the $Y$-space.
We need not solve these equations. From the previous results
we know that the following Yang-Mills-type functions are gauge invariant:
\begin{equation}\label{8YangMills}
  \left\{
    \begin{aligned}
      & A_{\text{YM}}^{+++}\equiv X_1^+ Y_1^{++}+X_2^+Y_2^{++}+X_3^+Y_3^{++}\\
      & A_{\text{YM}}^{++-}\equiv X_1^+Y_1^{+-}+X_2^+Y_2^{-+}+X_3^-Y_3^{++}\\
      & A_{\text{YM}}^{--+}\equiv X_1^-Y_1^{-+}+X_2^-Y_2^{+-}+X_3^+Y_3^{--}\\
      & A_{\text{YM}}^{+-+}\equiv X_1^+ Y_1^{-+}+X_2^- Y_2^{++}+X_3^+ Y_3^{+-}\\
      & A_{\text{YM}}^{-+-}\equiv X_1^- Y_1^{+-}+X_2^+ Y_2^{--}+X_3^- Y_3^{-+}\\
      & A_{\text{YM}}^{-++}\equiv X_1^- Y_1^{++}+X_2^+ Y_2^{+-}+X_3^+ Y_3^{-+}\\
      & A_{\text{YM}}^{+--}\equiv X_1^+ Y_1^{--}+X_2^- Y_2^{-+}+X_3^- Y_3^{+-}\\
      & A_{\text{YM}}^{---}\equiv X_1^-Y_1^{--}+X_2^-Y_2^{--}+X_3^-Y_3^{--}
    \end{aligned}~,
  \right.
\end{equation}
 and 7 of which  are  linearly independent.
We will choose the first 7 functions to be independent 
in~\ref{solution_MixedSymmetry}. 
A general solution can thus  be written as
\begin{equation}   \label{solution_MixedSymmetry}
    A=f_X(A_{\text{YM}}^{+++}, A_{\text{YM}}^{++-}, A_{\text{YM}}^{--+},
    A_{\text{YM}}^{+-+},A_{\text{YM}}^{-+-},A_{\text{YM}}^{-++},\ym{+--})~.
\end{equation}

\begin{figure}[htb]
    \begin{center}
        \includegraphics[width=8cm]{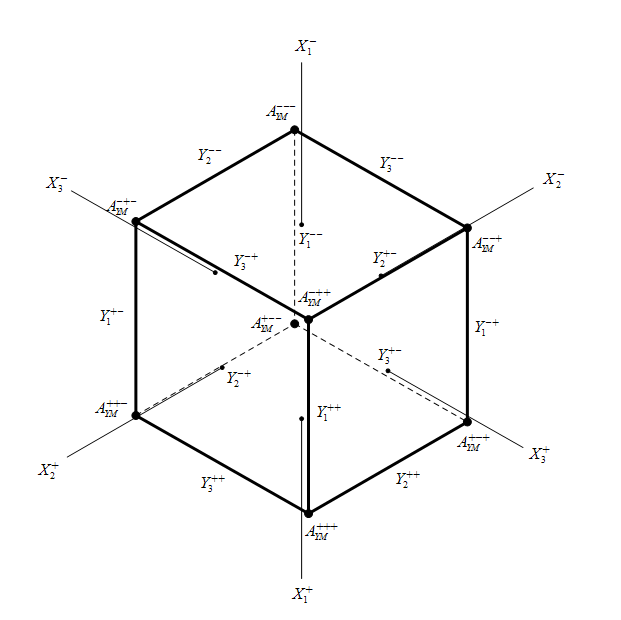}
        \caption{The relations (\ref{8YangMills}) represented by a cuboid, in which 
         the $X$'s are represented by the faces of the cuboid, the  $Y$'s are 
         represented by the edges,  and the functions 
         $A_{\text{YM}}^{\chi_1\chi_2\chi_3}$ 
         ($\chi_i=+$ or $-$) are represented by the vertices. Each 
         $A_{\text{YM}}^{\chi_1\chi_2\chi_3}$ is associated with three faces and
          three edges to which the corresponding vertex is attached. Each vertex  
          is  a sum of the associated faces multiplied by perpendicular edges.}
        \label{CuboidRepresentation}
    \end{center}
\end{figure}

Although we have obtained the solution~(\ref{solution_MixedSymmetry}),
we still need to impose the condition that
$f_X(A_{\text{YM}}^{\chi_1\chi_2\chi_3})$
 be a polynomial in  $X$'s and $Y$'s.
We can set 5 of the $Y$'s to zero such that the
7 of the $A_{\text{YM}}^{\chi_1\chi_2\chi_3}$
are still linearly independent.
In this way the remaining $Y$'s can be expressed as linear combinations of the
$A_{\text{YM}}^{\chi_1\chi_2\chi_3}$,
and
$f_X(A_{\text{YM}}^{\chi_1\chi_2\chi_3})$
must be a polynomial in
$A_{\text{YM}}^{\chi_1\chi_2\chi_3}$,
with rational functions of $X$'s as coefficients.
Furthermore  the amplitude is homogeneous in $e_i^{\pm}$,
these rational coefficient functions must therefore be homogeneous in
$X_i^\pm$ and must  be of the form
\begin{equation}
    \prod_{i=1}^3 (X_i^+)^{n_i^+} (X_i^-)^{n_i^-}~,
\end{equation}
where $n_i^\pm$ being   integers.
The amplitude can  finally be expressed as
\begin{equation}
 \label{general_amplitude}
  A=P\left(X,A_{\text{YM}}^{\chi_1\chi_2\chi_3}\right)
    +\frac{Q\left(X,A_{\text{YM}}^{\chi_1\chi_2\chi_3}\right)}
     {\displaystyle\prod_{i=1}^3 (X_i^+)^{m_{i}^+} (X_i^-)^{m_i^-}}
\end{equation}
where $m_i^+,m_i^-$ are non-negative integers and
$P,Q$ are polynomials in the $X$s and
$A_{\text{YM}}^{\chi_1\chi_2\chi_3}$.
The polynomial
$Q(X,A_{\text{YM}}^{\chi_1\chi_2\chi_3})$
is required not to contain a factor of $X_i^\pm$.

Let us consider the case where the only non-vanishing $m$ in~(\ref{general_amplitude}) is $m_3^+=1$:
\begin{equation}
    A_3^+=P_3^+\left(X,A_{\text{YM}}^{\chi_1\chi_2\chi_3}\right)
    +\frac{Q_3^+\left(X,A_{\text{YM}}^{\chi_1\chi_2\chi_3}\right)}{X_3^+}
\end{equation}
We note  the following identity:
$$
X_1^-X_2^-{A}_{\text{YM}}^{+++}+X_1^+X_2^+{A}_{\text{YM}}^{--+}
   -X_1^-X_2^+{A}_{\text{YM}}^{+-+}-X_1^+X_2^-{A}_{\text{YM}}^{-++}
    \equiv 0 \quad(\text{mod } X_3^+)
$$
which means that the LHS (denoted  by $K_3^+$)  contains a factor
$X_3^+$ when viewed as a polynomial in the $X$s and the $Y$s.
Expressing $A_{\text{YM}}^{-++}$ in terms of the other linearly independent
 $A_{\text{YM}}$'s and $K_3^+$:
$$
A_{\text{YM}}^{-++}=\frac{X_1^-X_2^-{A}_{\text{YM}}^{+++}
   +X_1^+X_2^+{A}_{\text{YM}}^{--+}-X_1^-X_2^+{A}_{\text{YM}}^{+-+}
    -K_3^+}{X_1^+X_2^+}
$$
we can rewrite $Q_3^+$ as
\begin{align*}
 & Q_3^+(X,A_{\text{YM}}^{+++},A_{\text{YM}}^{++-},A_{\text{YM}}^{--+},
  A_{\text{YM}}^{+-+},A_{\text{YM}}^{-+-},A_{\text{YM}}^{-++},\ym{+--})\\
=& \sum_{k=0}^{k_{\max}}(K_3^+)^k\widetilde{Q}_k(X,A_{\text{YM}}^{+++},
  A_{\text{YM}}^{++-},A_{\text{YM}}^{--+},A_{\text{YM}}^{+-+},
  A_{\text{YM}}^{-+-},\ym{+--})
\end{align*}
where $\widetilde{Q}_k$ are polynomials in
$A_{\text{YM}}^{+++}, \, A_{\text{YM}}^{++-}, \, A_{\text{YM}}^{--+}$,  and  
$A_{\text{YM}}^{+-+}, \, A_{\text{YM}}^{-+-}, \ym{+--}$~, 
which do not contain a factor $X_3^+$.
Let   ${A'}_{\text{YM}}^{\chi_1\chi_2\chi_3}
\equiv A_{\text{YM}}^{\chi_1\chi_2\chi_3}|_{X_3^+\rightarrow 0}$,
then
${A'}_{\text{YM}}^{+++}$,  ${A'}_{\text{YM}}^{++-}$, ${A'}_{\text{YM}}^{--+}$,
${A'}_{\text{YM}}^{+-+}$, ${A'}_{\text{YM}}^{-+-}$ and ${A'}_{\text{YM}}^{+--}$
are linearly independent. 
Therefore  if the coefficients of $\widetilde{Q}_0$
are not all zero, we would have
$$
    \widetilde{Q}_0\left(X\big|_{X_3^+\rightarrow 0},{A'}_{\text{YM}}^{+++},{A'}_{\text{YM}}^{++-},{A'}_{\text{YM}}^{--+},{A'}_{\text{YM}}^{+-+},{A'}_{\text{YM}}^{-+-},{A'}_{\text{YM}}^{+--}\right)\ne 0~.
$$
This indicates that
$\widetilde{Q}_0 \not \equiv 0\,\, (\text{mod }X_3^+)
\Rightarrow Q_3^+\not \equiv 0\,\,(\text{mod }X_3^+)$
which violates our assumption.
Hence the term $\widetilde{Q}_0$ must vanish and $Q_3^+$ contains a factor $K_3^+$.
So we have proved that $A_3^+$ can be written as
 \begin{equation}
    A_3^+=P_3^+\left(X,A_{\text{YM}}^{\chi_1\chi_2\chi_3}\right)+\beta_3^+ \widetilde{P}_3^+\left(X,A_{\text{YM}}^{\chi_1\chi_2\chi_3}\right)
 \end{equation}
where $\widetilde{P}_3^+$ is a polynomial in the $X$s and
$A_{\text{YM}}^{\chi_1\chi_2\chi_3}$,
and $\beta_3^+$ is defined by
\begin{align*}
    \beta_3^+&\equiv \frac{X_1^-X_2^-A_{\text{YM}}^{+++}+X_1^+X_2^+A_{\text{YM}}^{--+}-X_1^-X_2^+A_{\text{YM}}^{+-+}-X_1^+X_2^-A_{\text{YM}}^{-++}}{X_3^+}\\
    & = (e_1^+\cdot p_2)(e_2^+\cdot p_3)(e_1^-\cdot e_2^-)
         - (e_1^-\cdot p_2)(e_2^+\cdot p_3)(e_1^+\cdot e_2^-)\\
    &  - \,(e_1^+\cdot p_2)(e_2^-\cdot p_3)(e_1^-\cdot e_2^+)+ (e_1^-\cdot p_2)(e_2^-\cdot p_3)(e_1^+\cdot e_2^+)
\end{align*}

Likewise  we can show that a  generic amplitude~(\ref{general_amplitude})
can be cast into the following form,
\begin{equation}\label{form_of_amplitude}
    A=P\left(X,A_{\text{YM}}^{\chi_1\chi_2\chi_3}\right)+\prod_{i=1}^3(\beta_i^+)^{m_i^+}(\beta_i^-)^{m_i^-} \widetilde{P}\left(X,A_{\text{YM}}^{\chi_1\chi_2\chi_3}\right)
\end{equation}
where $m_i^\pm$ are non-negative integers,
$\widetilde{P}$ is a polynomial in the
$X$'s and $A_{\text{YM}}^{\chi_1\chi_2\chi_3}$, and
 \begin{equation}\label{form_beta}
 \left\{
 \begin{aligned}
    & \beta_1^\pm\equiv \frac{X_2^-X_3^-A_{\text{YM}}^{\pm++}-X_2^+X_3^-A_{\text{YM}}^{\pm-+}+X_2^+X_3^+A_{\text{YM}}^{\pm--}-X_2^-X_3^+A_{\text{YM}}^{\pm+-}}{X_1^\pm}\\
    & \beta_2^\pm\equiv \frac{X_1^-X_3^-A_{\text{YM}}^{+\pm+}-X_1^+X_3^-A_{\text{YM}}^{-\pm+}+X_1^+X_3^+A_{\text{YM}}^{-\pm-}-X_1^-X_3^+A_{\text{YM}}^{+\pm-}}{X_2^\pm}\\
    & \beta_3^\pm\equiv \frac{X_1^-X_2^-A_{\text{YM}}^{++\pm}-X_1^-X_2^+A_{\text{YM}}^{+-\pm}+X_1^+X_2^+A_{\text{YM}}^{--\pm}-X_1^+X_2^-A_{\text{YM}}^{-+\pm}}{X_3^\pm}\\
 \end{aligned}\right.
 \end{equation}
Staring at the figure~\ref{CuboidRepresentation},
the form of the other $\beta$'s can be easily inferred from $\beta_3^+$.
Since  $\beta_i^+=\beta_i^-$, we can set
$\beta_i\equiv \beta_i^+=\beta_i^-$.
We can conclude from~(\ref{form_of_amplitude})
 that the set of all gauge invariant 3-point amplitudes are thus  given by
\begin{equation}\label{set_of_amplitudes}
    A\in \mathbb{C}[A_{\text{YM}}^{\chi_1\chi_2\chi_3},X_i^\pm,\beta_i]~.
 \end{equation}
With $\beta_i$  being  antisymmetric in $e_j^+$ and $e_j^-$,  $j\ne i$. 
If  we restrict to the totally symmetric case~(\ref{polarizations}), the set (\ref{set_of_amplitudes}) reduces to
$\mathbb{C}\left[\ym{\chi_1\chi_2\chi_3},X_i^{\chi_i}\right]$,
with  fixed $\chi_1, \chi_2, \chi_3$.
That's why $\beta_i$ does not show up in equation~(\ref{amp_sss}).

\subsubsection{Helicity Amplitudes}

Now let us express the amplitudes in terms of  helicity spinors.
If there is no $\beta_i$ or self contraction terms $e^+_i\cdot e^-_i$, then an amplitude
$A[\epsilon_1(r_1,\sigma_1;\lambda_1),\epsilon_2(r_2,\sigma_2;\lambda_2),\epsilon_3(r_3,\sigma_3;\lambda_3)]$
can be expressed as a polynomial in
{$X_i^\pm$} and
$A_{\text{YM}}(e_1^\pm,e_2^\pm,e_3^\pm)$.
Adopting the convention in the literature~\cite{schwartz2014quantum},
\begin{gather*}
    p^{\alpha\dot{\alpha}}=p\rangle[p, \quad \quad 
    \left[\epsilon_p^-(r)\right]^{\alpha\dot{\alpha}}
    =\sqrt{2}\frac{p\rangle[r}{[pr]}, \quad \quad 
    \left[\epsilon_p^+(r)\right]^{\alpha\dot{\alpha}}
    =\sqrt{2}\frac{r\rangle[p}{\langle rp\rangle},
\end{gather*}
we have,
\begin{equation}
\label{amp_helicity}
\def\arraystretch{2.2}
    \begin{array}{lll}
  X_1^+\propto\dfrac{[12][31]}{[23]},
   & X_2^+\propto \dfrac{[12][23]}{[31]},
   & X_3^+\propto \dfrac{[31][23]}{[12]}\\
  X_1^-\propto\dfrac{\langle12\rangle\langle31\rangle}{\langle23\rangle},
   & X_2^-\propto \dfrac{\langle12\rangle\langle23\rangle}{\langle31\rangle},
   & X_3^-\propto \dfrac{\langle31\rangle\langle23\rangle}{\langle12\rangle}\\
  A_{\text{YM}}(e_1^+,e_2^+,e_3^-)\propto \dfrac{[12]^3}{[31][23]}
   & A_{\text{YM}}(e_1^+,e_2^-,e_3^+)\propto \dfrac{[31]^3}{[12][23]}
   & A_{\text{YM}}(e_1^-,e_2^+,e_3^+)\propto \dfrac{[23]^3}{[12][31]}\\
 A_{\text{YM}}(e_1^-,e_2^-,e_3^+)\propto
   \dfrac{\langle12\rangle^3}{\langle31\rangle\langle23\rangle}
   & A_{\text{YM}}(e_1^-,e_2^+,e_3^-)\propto
    \dfrac{\langle31\rangle^3}{\langle12\rangle\langle23\rangle}
   & A_{\text{YM}}(e_1^+,e_2^-,e_3^-)\propto
    \dfrac{\langle23\rangle^3}{\langle12\rangle\langle31\rangle}\\
    A_{\text{YM}}(e_1^+,e_2^+,e_3^+)=0
   & A_{\text{YM}}(e_1^-,e_2^-,e_3^-)=0 &
       \end{array}
\end{equation}
From~(\ref{amp_helicity}) we can see that a term in the polynomial
can be a product of only square brackets
(the first and the third row of~(\ref{amp_helicity})),
or a product of only angle brackets
(the second and the fourth row of~(\ref{amp_helicity})),
or a product of square brackets \textit{and} angle brackets.
The last type vanishes on-shell because the three on-shell momenta satisfy momentum conservation. 

The  amplitudes can, thus,  be written in the form
\begin{equation}
      \label{SHFResult}
\begin{aligned}
    & A[\epsilon_1(r_1,\sigma_1;\lambda_1),\epsilon_2(r_2,\sigma_2;\lambda_2),\epsilon_3(r_3,\sigma_3;\lambda_3)]\\
    = & \alpha_S {X_1^+}^{a_1^+}{X_2^+}^{a_2^+}{X_3^+}^{a_3^+}A_{\text{YM}}(e_1^+,e_2^+,e_3^-)^{b^{++-}}A_{\text{YM}}(e_1^+,e_2^-,e_3^+)^{b^{+-+}}A_{\text{YM}}(e_1^-,e_2^+,e_3^+)^{b^{-++}}\\
     &+ \alpha_A {X_1^-}^{a_1^-}{X_2^-}^{a_2^-}{X_3^-}^{a_3^-}
     A_{\text{YM}}(e_1^-,e_2^-,e_3^+)^{b^{--+}}
     A_{\text{YM}}(e_1^-,e_2^+,e_3^-)^{b^{-+-}}
     A_{\text{YM}}(e_1^+,e_2^-,e_3^-)^{b^{+--}}\\
    & + \text{VT...}\\
     = & \alpha'_S[12]^{\sigma_1+\sigma_2-\sigma_3}[23]^{\sigma_2+\sigma_3-\sigma_1}[31]^{\sigma_3+\sigma_1-\sigma_2}+\alpha'_A\langle12\rangle^{-\sigma_1-\sigma_2+\sigma_3}\langle23\rangle^{-\sigma_2-\sigma_3+\sigma_1}
     \langle31\rangle^{-\sigma_3-\sigma_1+\sigma_2}
\end{aligned}
\end{equation}
{where $\alpha_S,\alpha_A,\alpha'_S,\alpha'_A$ are numerical constants to be specified by the underlying theories, and}
\begin{align}
    &\left\{
        \begin{aligned}
            & b^{++-}=\frac{r_3-\sigma_3}{2},b^{+-+}=\frac{r_2-\sigma_2}{2},b^{-++}=\frac{r_1-\sigma_1}{2}\\
            & a_i^+=r_i-\frac{\sum_i r_i-\sum_i \sigma_i}{2}\qquad (i=1,2,3)
        \end{aligned}
    \right.\\
    &\left\{
        \begin{aligned}
            & b^{--+}=\frac{r_3+\sigma_3}{2},b^{-+-}=\frac{r_2+\sigma_2}{2},b^{+--}=\frac{r_1+\sigma_1}{2}\\
            & a_i^-=r_i-\frac{\sum_i r_i+\sum_i \sigma_i}{2}\qquad (i=1,2,3)
        \end{aligned}
    \right.
\end{align}
and $\{VT...\}$ stands for ``terms vanishing on-shell''.
 This coincides with the general form of three point amplitudes given by
  Benincasa and Cachazo~\cite{Benincasa:2007xk}.

The requirement, $a_1^+,a_2^+,a_3^+,b^{++-},b^{+-+},b^{-++}\geqslant 0$,  yields
\begin{equation}
\label{sigma_inequality1}
    \left\{
        \begin{aligned}
            & \sigma_i\leqslant r_i\quad (i=1,2,3)\\
            & \sum_i \sigma_i\geqslant \sum_i r_i-2\min (r_1,r_2,r_3)
        \end{aligned};
    \right.
\end{equation}
and  $a_1^-,a_2^-,a_3^-,b^{--+},b^{-+-}, b^{+--}\geqslant 0$ yields
\begin{equation}\label{sigma_inequality2}
    \left\{
        \begin{aligned}
            &\sigma_i\geqslant -r_i\quad (i=1,2,3)\\
            & \sum_i \sigma_i\leqslant -\left[\sum_i r_i-2\min (r_1,r_2,r_3)\right]
        \end{aligned}
    \right.
\end{equation}
The second inequalities in~(\ref{sigma_inequality1}) and (\ref{sigma_inequality2})
cannot  be simultaneously satisfied
(unless in the trivial case, $r_1=r_2=r_3=0$, which is not being  considered here).
This indicates that either $\alpha_S$ or $\alpha_A$ vanishes.
In the case where the coefficient $\alpha_A$ vanishes,
(\ref{sigma_inequality1}) leads    to
\begin{equation}
\label{square_nonvanishing_condition}
        \sum_i \sigma_i\geqslant \max(s_1,s_2,s_3) \quad (s_i\equiv |\sigma_i|)~.
\end{equation}
With $\sigma_1,\sigma_2,\sigma_3$ held fixed, this is a  necessary and sufficient
condition for the existence of non-vanishing $\alpha_S$:  if we choose 
$r_1=r_2=r_3=\max(s_1,s_2,s_3)$,
then the condition~(\ref{square_nonvanishing_condition}) implies ~(\ref{sigma_inequality1}).

Similarly, the necessary and sufficient condition for the existence of non-vanishing $\alpha_A$ is
\begin{equation}\label{angle_nonvanishing_condition}
        \sum_i \sigma_i\leqslant -\max(s_1,s_2,s_3)~.
\end{equation}
And  (\ref{square_nonvanishing_condition}) and (\ref{angle_nonvanishing_condition})
further indicate that  the signs of helicities
 can only be $+++$ or $++-$ for non-vanishing $\alpha_S$~
(In the latter case, the absolute value of the negative helicity should be less than or equal to the other two helicities.);  
for non-vanishing $\alpha_A$, the signs can only be $---$ or $--+$ 
(in the latter case, the positive helicity should be less than or equal to the absolute values of the other two helicities).

We pause here  to make a few  remarks.
  (\ref{sigma_inequality1}) and~(\ref{sigma_inequality2})
  imply that in the case of non-vanishing $\alpha_S$, the ranks of the polarization 
 tensors cannot exceed  $\sum \sigma_i$ and in the case of non-vanishing $\alpha_A$, 
 the ranks cannot  exceed $-\sum\sigma_i$.
  The inclusion of $\beta_i$ has no influence  on  our discussion since { $\beta_i$ vanishes in the spinor helicity formalism} and will only appear in the 
  $\{VT...\}$ terms.
Because a self contraction term $e_i^+\cdot e_i^-$ does not affect the helicity,
it will not affect~(\ref{square_nonvanishing_condition})
and~(\ref{angle_nonvanishing_condition}). 

{ 
The two  constraints in (\ref{square_nonvanishing_condition}) and (\ref{angle_nonvanishing_condition}) are absent in the light-cone approach. 
Thus, we reproduce the mismatch between the covariant approach and the light-cone 
approach. For further discussions on this mismatch, the readers are referred%
~\cite{Conde:2016izb, Bengtsson:2014qza, Ponomarev:2016lrm, Sleight:2016xqq}.
}

{ One can easily see from (\ref{SHFResult}) that the amplitudes 
are {independent of} $r_i$. This suggests that different representations with 
the same helicity can  give the same amplitudes, as long as the amplitudes for 
the  particular representation exist. But equation (\ref{SHFResult}) does not 
guarantee the existence of non-trivial amplitudes for the representations
 with mixed symmetries.}

\subsection{Amplitudes of Polarizations with Mixed Symmetry}\label{sec:amplitudes_mixed_symmetry}

We now turn our attention to scattering amplitudes involving tensors  with mixed symmetries upon permuting their Lorentz indices. 
We  denote $\bigotimes_{i=1}^{r} e(\mathbf{p},\pi(i))$ by $\hat{\epsilon}(r,\sigma;\pi)$ 
(with $\sum_{i=1}^r \pi(i)=\sigma$ and $\pi$ defined by~(\ref{polarization_expression})),  
and  denote a general polarization $\sum_\pi \lambda(\pi) \hat{\epsilon}(r,\sigma;\pi)$ by $\epsilon(r,\sigma;\lambda)$. 
Sometimes $\pi$ will also be used to denote the permutations that takes the 
canonically  ordered polarizations
$$\hat{\epsilon}^0(r,\sigma)\equiv e(\mathbf{p},+1)\otimes e(\mathbf{p},+1)\cdots e(\mathbf{p},-1)\otimes e(\mathbf{p},-1)$$
 into $\hat{\epsilon}(r,\sigma;\pi)$.

From (\ref{set_of_amplitudes}) we know that the amplitude
$A[\hat{\epsilon}_1(r_1,\sigma_1;\pi_1),\hat{\epsilon}_2(r_2,\sigma_2;\pi_2),\hat{\epsilon}_3(r_3,\sigma_3;\pi_3);N]$
can be written as a linear combination of the following expressions with 
 functions,  $\tau_i$  and $\theta_i$,  and integers,  $d_i$:
\begin{equation}
\label{amplitudes_hat}
\begin{aligned}
 & \prod_{k=1}^{\frac{r'_1-r'_2-r'_3+N'}{2}}e_1(\mathbf{p}_1,\tau_1(k))\cdot p_2\prod_{k=1}^{\frac{r'_2-r'_1-r'_3+N'}{2}}e_2(\mathbf{p}_2,\tau_2(k))\cdot p_1
    \prod_{k=1}^{\frac{r'_3-r'_1-r'_2+N'}{2}}e_3(\mathbf{p}_3,\tau_3(k))\cdot p_1\times\\
    & \prod_{k=1}^{\frac{r'_1+r'_2+r'_3-N'}{2}}A_{\text{YM}}\left[e_1(\mathbf{p}_1,\theta_1(k)),
    e_2(\mathbf{p}_2,\theta_2(k)),e_3(\mathbf{p}_3,\theta_3(k))\right]\cdot\beta_1^{d_1}\beta_2^{d_2}\beta_3^{d_3}
\end{aligned}
\end{equation}
where
\begin{align*}
    & 0\leqslant d_1+d_2\leqslant \frac{r_3-|\sigma_3|}{2},\,\,0\leqslant d_1+d_3\leqslant \frac{r_2-|\sigma_2|}{2},\,\,0\leqslant d_2+d_3\leqslant \frac{r_1-|\sigma_1|}{2}\\
    & r'_i\equiv r_i+2d_i-2\sum_{j=1}^3 d_j,\,\,N'\equiv N-2\sum_{j=1}^3 d_i
\end{align*}
 and $\pi_i,\theta_i$ being any functions 
\begin{align}
    & \tau_i: \left\{1,2,\cdots,\frac{2r'_i-(r'_1+r'_2+r'_3-N')}{2}\right\}\rightarrow \{+1,-1\}\\
    & \theta_i: \left\{1,2,\cdots,\frac{r'_1+r'_2+r'_3-N'}{2}\right\}\rightarrow \{+1,-1\}
\end{align}
that satisfy
$$\sum_{k=1}^{\frac{2r'_i-(r'_1+r'_2+r'_3-N')}{2}}
\hspace{-1.3em}\tau_i(k)\quad +\sum_{k=1}^{\frac{r'_1+r'_2+r'_3-N'}{2}}~.
\hspace{-1em}\theta_i(k)=\sigma_i~.
$$

In order to determine the gauge invariant amplitudes of general polarizations
$\epsilon_1(r_1,\sigma_1;\lambda_1)$, $\epsilon_2(r_2,\sigma_2;\lambda_2)$ 
and  $\epsilon_3(r_3,\sigma_3;\lambda_3)$,
we first write the amplitudes in the following forms,
\begin{equation}
\begin{aligned}
    A[\hat{\epsilon}_1(r_1,\sigma_1;\pi_1),\hat{\epsilon}_2(r_2,\sigma_2;\pi_2),\hat{\epsilon}_3(r_3,\sigma_3;\pi_3);N]
    &\equiv\hat{\epsilon}_1^{\mu_1\cdots\mu_{r_1}}\hat{\epsilon}_2^{\nu_1\cdots\nu_{r_2}}\hat{\epsilon}_3^{\rho_1\cdots\rho_{r_3}}
    \hat{K}_{\mu_1\cdots\mu_{r_1}\nu_1\cdots\nu_{r_2}\rho_1\cdots\rho_{r_3}}\\ 
    A[\epsilon_1(r_1,\sigma_1;\lambda_1),\epsilon_2(r_2,\sigma_2; \lambda_2),\,\epsilon_3(r_3,\sigma_3;\lambda_3);N]
    &\equiv \epsilon_1^{\mu_1\cdots\mu_{r_1}}\epsilon_2^{\nu_1\cdots\nu_{r_2}}\epsilon_3^{\rho_1\cdots\rho_{r_3}}K_{\mu_1\cdots\mu_{r_1}\nu_1\cdots\nu_{r_2}\rho_1\cdots\rho_{r_3}}
\end{aligned}\\
\end{equation}
then the gauge invariance conditions read
\begin{equation} \label{equation_K_hat}
\begin{aligned}
    \delta_i^{\pm}(\hat{\epsilon}_1^{\mu_1\cdots\mu_{r_1}}
    \hat{\epsilon}_2^{\nu_1\cdots\nu_{r_2}}
    \hat{\epsilon}_3^{\rho_1\cdots\rho_{r_3}})
    \hat{K}_{\mu_1\cdots\mu_{r_1}\nu_1
    \cdots\nu_{r_2}\rho_1\cdots\rho_{r_3}}=0
    \qquad(i=1,2,3)
    \end{aligned}
\end{equation}
\begin{equation} \label{equation_K}
   \delta_i^{\pm}(\epsilon_1^{\mu_1\cdots\mu_{r_1}}
    \epsilon_2^{\nu_1\cdots\nu_{r_2}}\epsilon_3^{\rho_1\cdots\rho_{r_3}})
  K_{\mu_1\cdots\mu_{r_1}\nu_1\cdots\nu_{r_2}\rho_1\cdots\rho_{r_3}}=0
  \qquad(i=1,2,3) 
\end{equation}
which are linear equations in $\hat{K}$ and $K$.

The solutions to equations~(\ref{equation_K_hat}) are given by the linear combinations of~(\ref{amplitudes_hat}) and equations~(\ref{equation_K}),  yielding
\begin{equation}
\begin{aligned}
    0=&\delta_i^{\pm}\left[\sum_{\pi_1}\lambda_1(\pi_1)
    \pi_1(\hat{\epsilon}^0_1)^{\mu_1\cdots\mu_{r_1}}
 \sum_{\pi_2}\lambda_2(\pi_2)\pi_2(\hat{\epsilon}^0_2)^{\nu_2\cdots\nu_{r_2}}
 \sum_{\pi_3}\lambda_3(\pi_3)\pi_3(\hat{\epsilon}^0_3)^{\rho_3\cdots\rho_{r_3}}\right]\\
 &\times K_{\mu_1\cdots\mu_{r_1}\nu_1\cdots\nu_{r_2}\rho_1\cdots\rho_{r_3}}\\
 = &\delta_i^{\pm}\left[(\hat{\epsilon}_1^0)^{\mu_1\cdots\mu_{r_1}}
   (\hat{\epsilon}_2^0)^{\nu_1\cdots\nu_{r_2}}
   (\hat{\epsilon}_3^0)^{\rho_1\cdots\rho_{r_3}}\right]
  \sum_{\pi_1}\sum_{\pi_2}\sum_{\pi_3}
  \bigg\{\lambda_1(\pi_1)\lambda_2(\pi_2)\lambda_3(\pi_3)\\
  &\times (\pi_1)^{-1}(\pi_2)^{-1}(\pi_3)^{-1}
   K_{\mu_1\cdots\mu_{r_1}\nu_1\cdots\nu_{r_2}\rho_1\cdots\rho_{r_3}}\bigg\}~.
\end{aligned}
\end{equation}

Comparing these equations with~(\ref{equation_K_hat})
we find that the solution space for $K$ can be obtained by
acting with the linear operators
$\sum_{\pi_1}\sum_{\pi_2}\sum_{\pi_3}\lambda_1(\pi_1)\lambda_2(\pi_2)\lambda_2(\pi_2)(\pi_1)^{-1}(\pi_2)^{-1}(\pi_3)^{-1}$
(which can be  singular) on the solution space of~(\ref{equation_K_hat}).
We  thus have
\begin{equation}
\label{amplitude_particular}
\begin{aligned}
  & A\left[\epsilon_1(r_1,\sigma_1;\lambda_1),\epsilon_2(r_2,\sigma_2;\lambda_2),\epsilon_3(r_3,\sigma_3;\lambda_3);N\right]\\
 = & \epsilon_1^{\mu_1\cdots\mu_{r_1}}\epsilon_2^{\nu_1\cdots\nu_{r_2}}\epsilon_3^{\rho_1\cdots\rho_{r_3}}\hat{K}_{\mu_1\cdots\mu_{r_1}\nu_1\cdots\nu_{r_2}\rho_1\cdots\rho_{r_3}}\\
 = &\sum_{\pi_1}\sum_{\pi_2}\sum_{\pi_3}
 \lambda_1(\pi_1)\lambda_2(\pi_2)\lambda_3(\pi_3)
  A[\hat{\epsilon}_1(r_1,\sigma_1;\pi_1),\hat{\epsilon}_2(r_2,\sigma_2;\pi_2),\hat{\epsilon}_3(r_3,\sigma_3;\pi_3);N]~.
\end{aligned}
\end{equation}

\section{Examples}
\label{sec:examples}

In this section we present a few concrete examples to illustrate our  procedure of
constructing gauge invariant amplitudes.
 With  the polarizations denoted by
$\epsilon_1,\epsilon_2,\epsilon_3$,  the amplitudes can be classified into
different categories labelled by ($\overline{L},\overline{M},\overline{N}$),
where $\overline{L}$ denotes the number of Lorentz contractions, 
$e_i^+\cdot e_i^-=\text{const}$, 
$\overline{M}$ the number of $\beta_i$ and
 $\overline{N}$ the number of $\ym{\chi_1\chi_2\chi_3}$
in each term of a given  amplitude.
This classification includes all possible amplitudes, 
but those of different types 
may not be linearly independent (as we will discuss further  below).
The number of derivatives  (denoted by $N$) in each term of an
amplitude of type ($\overline{L},\overline{M},\overline{N}$) is given by 
\begin{equation}
    N=r_1+r_2+r_3-2(\overline{L}+\overline{M}+\overline{N})
\end{equation}
where $r_1,r_2,r_3$ are the ranks of the polarizations.

Let us first consider the case
\begin{equation}
    \epsilon_1=e_1^+\otimes e_1^-,\epsilon_2=e_2^+\otimes e_2^-,\epsilon_3=e_3^+\otimes e_3^-~.
\end{equation}
Amplitudes  $A(\epsilon_1,\epsilon_2,\epsilon_3)$ are 
 thus (where $C_i\equiv e_i^+\cdot e_i^-$) classified:
\begin{center}
$\def\arraystretch{1.5}
\begin{array}{|c|c|}
    \hline
    \text{Categories} & \text{Amplitudes}\\ \hline
    (0,0,0) & X_1^+X_1^-X_2^+X_2^-X_3^+X_3^-\\ \hline
    (0,0,1) &
    \begin{aligned}
   & X_1^- X_2^- X_3^-\ym{+++},\quad X_1^+X_2^-X_3^-\ym{-++},
   \quad X_1^-X_2^+X_3^-\ym{+-+},\quad X_1^-X_2^-X_3^+\ym{++-}\\
        & X_1^-X_2^+X_3^+\ym{+--},\quad X_1^+X_2^-X_3^+\ym{-+-},
        \quad X_1^+X_2^+X_3^-\ym{--+}
    \end{aligned}
    \\
    \hline
    (0,0,2) & \ym{+++}\ym{---},\quad \ym{-++}\ym{+--},
    \quad\ym{+-+}\ym{-+-},\quad\ym{++-}\ym{--+}\\ \hline
    (0,1,0) & \beta_1X_1^+X_1^-,\quad \beta_2X_2^+X_2^-,
    \quad \beta_3X_3^+X_3^-
    \\
    \hline
    (1,0,0) & C_1X_2^+X_2^-X_3^+X_3^-,\quad C_2X_3^+X_3^-X_1^+X_1^-,
    \quad C_3X_1^+X_1^-X_2^+X_2^-\\
    \hline
    (1,1,0) & C_1\beta_1,\quad C_2\beta_2,\quad C_3\beta_3\\
    \hline
    (2,0,0) & C_2C_3X_1^+X_1^-,\quad  C_1C_3X_2^+X_2^-,
    \quad  C_1C_2X_3^+X_3^-\\
    \hline
    (3,0,0) & C_1C_2C_3\\
    \hline
\end{array}
$\captionof{table}{List$(+-,+-,+-)$}
\end{center}
We refer to this list as List$(+-,+-,+-)$.

One may attempt to include another amplitude $X_1^+X_2^+X_3^+\ym{---}$ of type (0,0,1) in the above list but,  as we have mentioned before, it has linear dependence on the other seven amplitudes of the same type, 
namely,  
\begin{align*}
    X_1^+X_2^+X_3^+\ym{---}& = X_1^- X_2^- X_3^-\ym{+++}-X_1^+X_2^-X_3^-\ym{-++}-X_1^-X_2^+X_3^-\ym{+-+}-X_1^-X_2^-X_3^+\ym{++-}\\ & +X_1^-X_2^+X_3^+\ym{+--}+X_1^+X_2^-X_3^+\ym{-+-}+X_1^+X_2^+X_3^-\ym{--+}
\end{align*}

Categories that have a common number $\overline{L}+\overline{M}+\overline{N}$ 
(or in other words, have the same number of derivatives) 
 contain linearly dependent amplitudes. 
This happens when amplitudes with a term $\beta_i X_i^{\pm}$ appear in the list:
 $\beta_i X_i^{\pm}$ is a polynomial in $X$s and $\ym{\chi_1\chi_2\chi_3}$.
Therefore  we need to remove all the amplitudes of type (0,1,0).

Let us now determine the gauge invariant amplitudes for the following cases:
\begin{equation}\label{ex_polarizations}
\begin{aligned}
    (1)\,\,\, &\epsilon_1=e_1^+\otimes e_1^--e_1^-\otimes e_1^+,\quad \epsilon_2=e_2^{\chi_2}\otimes e_2^{\chi_2},\quad \epsilon_3=e_3^{\chi_3}\otimes e_3^{\chi_3}\\
    (2)\,\,\, &\epsilon_1=e_1^+\otimes e_1^--e_1^-\otimes e_1^+,\quad \epsilon_2=e_2^+\otimes e_2^--e_2^-\otimes e_2^+,\quad \epsilon_3=e_3^{\chi_3}\otimes e_3^{\chi_3}\\
    (3)\,\,\, &\epsilon_1=e_1^+\otimes e_1^--e_1^-\otimes e_1^+,\quad \epsilon_2=e_2^+\otimes e_2^--e_2^-\otimes e_2^+,\quad \epsilon_3=e_3^+\otimes e_3^--e_3^-\otimes e_3^+~.
\end{aligned}
\end{equation}

In Case~(1), we substitute $e_2^+\otimes e_2^-$, in List$(+-,+-,+-)$,  
 by $e_2^{\chi_2}\otimes e_2^{\chi_2}$ and $e_3^+\otimes e_3^-$ by
$e_3^{\chi_3}\otimes e_3^{\chi_3}$ to obtain  List$(+-,\chi_2\chi_2,\chi_3\chi_3)$, 
 where in each amplitude:
$$
\#(e_1^+)=\#(e_1^-)=1,\#(e_2^{\chi_2})=\#(e_3^{\chi_3})=2,\#(e_2^{\bar{\chi}_2})=\#(e_3^{\bar{\chi}_3})=0~.
$$
$\chi_i$ can be either $+$ or $-$,  
and $\bar{\chi}_i$ denotes the sign opposite to $\chi_i$. 

We then delete from List$(+-,\chi_2\chi_2,\chi_3\chi_3)$ the amplitudes that are symmetric in $e_1^+$ and $e_1^-$ to get the following truncated list,
\begin{center}
$\def \arraystretch{1.5}
\begin{array}{|c|c|}
    \hline
    \text{Categories} & \text{Amplitudes}\\ \hline
    (0,0,1) & X_1^- X_2^{\chi_2} X_3^{\chi_3}\ym{+\chi_2\chi_3},\quad X_1^+X_2^{\chi_2} X_3^{\chi_2}\ym{-\chi_2\chi_3}\\ \hline
\end{array}
$\captionof{table}{Truncation of List$(+-,\chi_2\chi_2,\chi_3\chi_3)$}
\end{center}

A general gauge invariant amplitude of the polarizations in (1) can be obtained
by anti-symmetrizing  the amplitudes in the list over the indices of $\epsilon_1$:
\begin{equation}\label{A1}
\begin{aligned}
    A^{(1)}=& X_1^- X_2^{\chi_2} X_3^{\chi_3}\ym{+\chi_2\chi_3}-X_1^+X_2^{\chi_2} X_3^{\chi_3}\ym{-\chi_2\chi_3}\\
    = &(p_3\cdot\epsilon_2\cdot p_3)(p_1\cdot\epsilon_3\cdot\epsilon_1\cdot p_2)+(p_2\cdot \epsilon_3\cdot p_2)(p_1\cdot \epsilon_2\cdot\epsilon_1\cdot p_3)
\end{aligned}
\end{equation}

Similarly, the gauge invariant amplitudes for Case~(2) can be obtained by
anti-symmetrizing the amplitudes in the following list over the indices of
$\epsilon_1$ and then over the indices of $\epsilon_2$,
\begin{center}
$\def \arraystretch{1.5}
\begin{array}{|c|c|}
    \hline
    \text{Categories} & \text{Amplitudes}\\ \hline
    (0,0,1) &
    \begin{aligned}
        & X_1^- X_2^- X_3^{\chi_3}\ym{++\chi_3},\,\, X_1^+X_2^-X_3^{\chi_3}\ym{-+\chi_3},\,\, X_1^-X_2^+X_3^{\chi_3}\ym{+-\chi_3},\,\, X_1^+X_2^+X_3^{\chi_3}\ym{--\chi_3}
    \end{aligned}\\ \hline
    (0,0,2) & \ym{++\chi_3}\ym{--\chi_3},\quad \ym{-+\chi_3}\ym{+-\chi_3}\\ \hline
\end{array}
$\captionof{table}{Truncation of List$(+-,+-,\chi_3\chi_3)$}
\end{center}

The result is:
\begin{equation}\label{A2}
\left\{\begin{aligned}
&\begin{aligned}
    A^{(2)}_{\,\,1}=&  X_1^+X_2^+X_3^{\chi_3}\ym{--\chi_3}-X_1^-X_2^+X_3^{\chi_3}\ym{+-\chi_3}-X_1^+X_2^-X_3^{\chi_3}\ym{-+\chi_3}+X_1^-X_2^-X_3^{\chi_3}\ym{++\chi_3}\\
    =& (p_1\cdot \epsilon_3\cdot p_1)(p_2\cdot \epsilon_1\cdot\epsilon_2 \cdot p_3),
\end{aligned}\\
&\begin{aligned}
    A^{(2)}_{\,\,2}=& \ym{++\chi_3}\ym{--\chi_3}-\ym{+-\chi_3}\ym{-+\chi_3}\\
    =& 2(p_2\cdot\epsilon_1\cdot\epsilon_3\cdot\epsilon_2\cdot p_3-p_1\cdot \epsilon_3\cdot \epsilon_2\cdot\epsilon_1\cdot p_2-p_1\cdot \epsilon_3\cdot \epsilon_1\cdot\epsilon_2\cdot p_3)+(\epsilon_1^{\mu\nu}\epsilon_{2\mu\nu})(p_1\cdot\epsilon_3\cdot p_1)
\end{aligned}
\end{aligned}\right.
\end{equation}

Furthermore  the amplitudes for Case~(3) (if exists)  can be obtained from the 
following list. But as one can check, the amplitudes listed below all vanish 
upon anti-symmetrization over the indices of $\epsilon_1$, $\epsilon_2$ and $\epsilon_3$.
\begin{center}
$\def \arraystretch{1.5}
\begin{array}{|c|c|}
    \hline
    \text{Categories} & \text{Amplitudes}\\ \hline
    (0,0,1) &
    \begin{aligned}
        & X_1^- X_2^- X_3^-\ym{+++},\quad X_1^+X_2^-X_3^-\ym{-++},\quad X_1^-X_2^+X_3^-\ym{+-+},\quad X_1^-X_2^-X_3^+\ym{++-}\\
        & X_1^-X_2^+X_3^+\ym{+--},\quad X_1^+X_2^-X_3^+\ym{-+-},\quad X_1^+X_2^+X_3^-\ym{--+}
    \end{aligned}\\ \hline
    (0,0,2) & \ym{+++}\ym{---},\quad \ym{-++}\ym{+--},\quad\ym{+-+}\ym{-+-},\quad\ym{++-}\ym{--+}\\ \hline
\end{array}
$\captionof{table}{Truncation of List$(+-,+-,+-)$}
\end{center}

{ Although in four dimensions an anti-symmetric rank-2 tensor field $A_{\mu\nu}$ 
is dual to the scalar field $\phi$ in the free theory, as they both describe 
a spin-0 degree of freedom, they are no longer dual to each other once interaction is introduced.  
Whereas non-trivial 3-point amplitudes for $A_{\mu\nu}$ 
self interaction is absent,  $\phi^3$ can exist and 
lead to non-trivial amplitudes. 
This is the result of different gauge transformations for these two fields. 
As a matter of fact we  can view the polarization of the scalar field $\phi$ as 
$1$  as it does not have any gauge transformations. 
We could write down an amplitude $1^3$ for $\phi^3$
(up to a coupling constant). 
For  $A_{\mu\nu}$, its   polarization tensor is   
$\epsilon_1=e_1^+\otimes e_1^--e_1^-\otimes e_1^+$
in the equivalence  class 
  of the 2-dim  Levi-Civita tensor. 
One  could contract three  Levi-Civita tensors to obtain
 a non-zero result which 
could potentially be the  amplitudes dual to  $\phi^3$. 
But when we express the contraction in terms of 
$\epsilon_{1\mu\nu}$ 
$\epsilon_{2\mu\nu}$ $\epsilon_{3\mu\nu}$~\footnote{
Since $\phi^3$ is not a derivative coupling, the dual amplitude should  not contain momentum.}, 
none   are  gauge invariant.  
 Even if we relax the constraint of parity conservation, 
 allowing the 4-d Levi-Civita tensor $\epsilon_{\mu\nu\rho\sigma}$ to appear 
 in the amplitudes, non-trivial gauge invariant 
 3-point amplitudes describing  $A_{\mu\nu}$ self-interaction do not exist.

If we, however, only consider  expressions  with many $\epsilon_{\mu\nu}$'s 
which are polarization tensors for the same momentum $p$, then as long as we contract 
all indices to obtain scalars, the resultant expressions are always gauge invariant.  
But this case only appears in the free theory 
and is not true for an interactive theory since with interaction  we always 
need to construct scalars from  different $\epsilon_{i\mu\nu}$ corresponding to 
momenta  $p_{i}$. 
This is yet another way to see that the duality between rank-2 antisymmetric tensor and scalar fields does not extend to an interactive theory. }

Having determined the gauge invariant 3-point amplitudes in the above examples,
we shall  find the cubic interaction terms from which these amplitudes can be derived.
Here we assume that particles with the same kind of polarization tensors are 
identical and that the polarizations $\epsilon_1,\epsilon_2,\epsilon_3$ in (\ref{ex_polarizations}) correspond to the fields $\phi_a$,$\phi_b$,$\phi_c$
which satisfy the Lorenz gauge condition:
\begin{equation}\label{LorenzCondition}
	\partial^{\mu}\phi_{k\mu\nu} =\partial^{\nu}\phi_{k\mu\nu} = 0\quad (k=a,b,c)
\end{equation}
This gauge condition is implicitly imposed by equation~(\ref{tensorConditionModified}). 
Carrying out the following replacement (where the subscript $i$ of $(\partial_\mu)_i$ indicates which field the partial derivative acts upon),
\begin{equation}\label{replacement}
\def \arraystretch{1.5}\setlength\arraycolsep{15pt}
    \begin{array}{lll}
        \epsilon_{1\mu\nu}\rightarrow \phi_{a\mu\nu} & \epsilon_{2\mu\nu}\rightarrow \phi_{b\mu\nu} & \epsilon_{3\mu\nu}\rightarrow \phi_{c\mu\nu}\\
        \,\,p_{\mu}\,\,\rightarrow (\partial_\mu)_a &\, \,q_{\mu}\,\,\rightarrow (\partial_\mu)_b &\,\,k_{\mu}\,\,\rightarrow (\partial_\mu)_c
    \end{array}
\end{equation}
we obtain the interaction terms tabulated below,
$$
\def\arraystretch{1.7}
\begin{array}{|c|c|}
\hline
    \text{amplitude} & \text{interaction term}\\ 
\hline
     A^{(1)}
    & f^{abc}\left[(\partial^\kappa\phi_b^{\mu\nu})(\partial_\nu \partial_\mu\phi_c^{\rho\sigma})(\partial_\rho\phi_{a\sigma\kappa})+(\partial^\kappa\phi_c^{\mu\nu})(\partial_\nu\partial_\mu\phi_b^{\rho\sigma})(\partial_\rho\phi_a^{\sigma\kappa})\right]\\ 
\hline
     A^{(2)}_{\,\,1} & f^{abc}(\partial_\mu\partial_\nu\phi_a^{\rho\sigma})(\partial_\rho\phi_{b\sigma\kappa})(\partial^\kappa\phi_c^{\mu\nu})\\ 
 \hline
    A^{(2)}_{\,\,2}
    & \begin{aligned}
     f^{abc} &\Big[2\phi_{a\mu\nu}(\partial^\mu \phi_{b\rho\sigma})(\partial^\sigma \phi_c^{\nu\rho})-2\phi_{c\mu\nu}(\partial^\mu \phi_{a\rho\sigma})(\partial^\sigma \phi_b^{\nu\rho})-2\phi_{b\rho\sigma}(\partial^\mu \phi_a^{\nu\rho})(\partial^\sigma \phi_{c\mu\nu})\\
    & +(\partial_\rho\partial_\sigma \phi_a^{\mu\nu})\phi_{b\mu\nu}\phi_c^{\rho\sigma}\Big]
    \end{aligned}\\ 
\hline
\end{array}
$$
Because of the gauge condition (\ref{LorenzCondition}),
the form of these interaction terms is not unique -- a term
$$
(\cdots\phi_a^{\cdots\mu\cdots}\cdots\partial_\mu\phi_b^{\cdots}\cdots\phi_c^{\cdots})
$$
can be changed into
$$
-(\cdots\phi_a^{\cdots\mu\cdots}\cdots\phi_b^{\cdots}\cdots\partial_\mu\phi_c^{\cdots})
$$
by integration by parts. This corresponds to the fact that
$p_{2\mu}\epsilon_1^{\cdots\mu\cdots}=-p_{3\mu}\epsilon_1^{\cdots\mu\cdots}$.

{Our previous discussions are all based on on-shell gauge invariance. If we go off-shell, then further information about the interactions can be obtained. For example, we have obtained from on-shell gauge invariance the Yang-Mills amplitude, which allows us to determine the Yang-Mills Lagrangian up to cubic terms:
\begin{equation}
	\mathcal{L}=-\frac{1}{4}(\partial_\mu A_\nu^a-\partial_\nu A_\mu^a)(\partial^\mu A^{a\nu}-\partial^\nu A^{a\mu})-gf^{abc}(\partial_\mu A^a_\nu)A^{b\mu}A^{c\nu}
\end{equation}
with $f^{abc}$ being antisymmetric. Under the off-shell version of the leading order gauge transformation we have mentioned, the second term in the Lagrangian will not be invariant:
\begin{equation}\label{eq:YMgaugeT}
\begin{aligned}
	&\delta\left[-gf^{abc}(\partial_\mu A^a_\nu)A^{b\mu}A^{c\nu}\right]\\
	\rightarrow & gf^{abc}\left[2\xi^c\partial_\mu\partial_\nu(A^{b\mu}A^{a\nu})+\xi^c\partial^\mu(\partial_\mu A_\nu^b A^{a\nu})-\xi^c\partial_\nu(\partial_\mu A^{b\mu}A^{a\nu})\right]\\
	=&gf^{abc}\xi^c(\partial^2 A_\nu^b)A^{a\nu}
\end{aligned}
\end{equation}
where the right arrow indicates that we have performed integration by parts.
To compensate for this term, we have to add in the gauge transformation an extra term to make the change of the kinetic term first order in $g$. This extra term must be bilinear in $A_\mu^a$ and $\xi^a$:
\begin{equation}
	\delta A_\mu^a=\partial_\mu \xi^a+g F^{abc}A_\mu^b \xi^c
\end{equation}
where $F^{abc}$ are constants to be determined. Then the variation of the kinetic term becomes
\begin{equation}\label{eq:KgaugeT}
\begin{aligned}
	&\delta\left[-\frac{1}{4}(\partial_\mu A_\nu^a-\partial_\nu A_\mu^a)(\partial^\mu A^{a\nu}-\partial^\nu A^{a\mu})\right]\\
	\rightarrow &gF^{abc}\xi^c\left\{\partial^2 A^{a\nu}A_\nu^b-\left[\partial_\mu\partial_\nu A^{a\mu}A^{b\nu}+(\mu\leftrightarrow\nu)]\right]\right\}
\end{aligned}
\end{equation}
Now that (\ref{eq:YMgaugeT})+(\ref{eq:KgaugeT})=0, we have $F^{abc}=f^{abc}$. We only focus on the on-shell case and will not explore this further. For more discussions on this topic, see e.g.~\cite{Berends:1984rq}.
}

\section{Conclusion and Discussion}
\label{sec:conclusion}
We found that under the assumptions of locality, Lorentz invariance and parity conservation, a general three-point  amplitude of
massless higher-spin gauge bosons can be written as a polynomial in
$\ym{\chi_1\chi_2\chi_3}$, $X_{i}=e_{i}\cdot p_{j}^\pm$,  
$\beta_i$~(as defined by~(\ref{form_beta})),
and self contraction terms $e_i^+\cdot e_i^-$. 
 For a quantum field theory in four dimensional flat spacetime --~either renormalizable 
 or non-renormalizable--the interaction terms are narrowed down  (at least in cubic vertices) by the Lorentz invariance and locality properties to  only  a few choices. 
  This  is not only a perturbative, but also a non-perturbative constraint. 

If the polarizations are totally symmetric, then the $\beta_i$ terms vanish.
For three particular polarizations, the possible amplitudes are determined by equation~(\ref{amplitude_particular}) as illustrated in Section~\ref{sec:examples}.
 We also computed explicitly  the helicity amplitudes to show
 that the helicities must satisfy~(\ref{square_nonvanishing_condition}) 
 or~(\ref{angle_nonvanishing_condition})  for the amplitudes to be nontrivial.

The amplitudes for totally symmetric polarizations
$\epsilon^{\pm}_i(p)=\bigotimes_{i=1}^{s}e^{\pm}_i(p)$ are given by
\begin{equation}
    \begin{aligned}
  & A(\epsilon_1,\epsilon_2,\epsilon_3;N)\\
  = & (e_1\cdot p_2)^{\frac{s_1-s_2-s_3+N}{2}}(e_2\cdot p_1)^{\frac{s_2-s_1-s_3+N}{2}}(e_3\cdot p_1)^{\frac{s_3-s_1-s_2+N}{2}}A_{\text{YM}}(e_1,e_2,e_3)^{\frac{s_1+s_2+s_3-N}{2}}
    \end{aligned}
\end{equation}
where $N$ is the total number of derivatives in the cubic vertex and
\begin{equation}
A_{\text{YM}}(e_1,e_2,e_3)\equiv (e_1\cdot p_2)(e_2\cdot e_3)
-(e_2\cdot p_1)(e_1\cdot e_3)+(e_3\cdot p_1)(e_1\cdot e_2)
\end{equation}
In 4-dimension, however, due to a Schouten-like identity, the non-trivial amplitudes are given by (assuming $s_1\leq s_2\leq s_3$ for convenience)
\begin{equation}
	(e_1\cdot p_2)^{s_1} (e_2\cdot p_1)^{s_2} (e_3\cdot p_1)^{s_3}, \quad (e_2\cdot p_1)^{s_2 - s_1} (e_3\cdot p_1)^{s_3 - s_1} A_{YM}^{s_1}
\end{equation}

The amplitudes involving the second rank totally symmetric polarizations $e_i\otimes e_i\,\,(i=1,2,3)$ are of particular interest, because they are related to gravity.
\begin{equation*}
A_{\text{YM}}^2(e_1,e_2,e_3), \quad (e_1\cdot p_2)^2(e_2\cdot p_3)^2(e_3\cdot p_1)^2
\end{equation*}
with momentum number $N=2$, $N=6$, respectively. 
A few remarks are due: 
\begin{itemize}
 \item
 In the case of  $N=2$, the amplitude can be written as a product of two Yang-Mills
  amplitudes, which is consistent with the  Einstein-Hilbert action.
 It was shown by Boels and Medina~\cite{Boels:2016xhc} that higher-point Einstein-Hilbert amplitudes can be obtained from the 3-point amplitude by imposing the symmetry
 and unitarity conditions. Furthermore this is  the only amplitude (among the two possible ones that we found here) that corresponds to a constructible theory~\cite{Benincasa:2007xk}, and as predicted by  the famous KLT relation between open and closed strings~\cite{Kawai:1985xq}.
\item
In the case of  $N=6$, the amplitude--similar to  the Einstein-Hilbert case--is 
symmetric under permutations of the particles, thus the particle can be a singlet.
\end{itemize}
	
Similar to the case of  rank-2 tensors, the possible amplitudes of three totally symmetric
polarizations of rank-$r$ tensors are given by
\begin{equation}\label{final}
A_{\text{YM}}^r(e_1,e_2,e_3), \quad (e_1\cdot p_2)^r(e_2\cdot p_3)^r(e_3\cdot p_1)^r
\end{equation}
When $r$ is even, (\ref{final}) is symmetric under permutations of the particles and the particles in the theory can be a singlet; when $r$ is odd, (\ref{final}) is antisymmetric and the theory is Yang-Mills like, i.e. carrying color indices.

In 4-dimension,  massless mixed-symmetry fields can be  dualized  to  totally symmetric fields in the free theory. Our analysis shows that this 
duality does not extend to the interacting theory. 
There exists a mismatch between the 3-point amplitudes and thus  the cubic interactions of these two fields cannot be dual to each other.  
To be more precise, in the case of the antisymmetric rank-2 field $A_{\mu\nu}$, 
although it is dual to the scalar field $\phi$ which enjoys a non-trivial 3-point amplitude, we cannot find a non-trivial 3-point amplitude for $A_{\mu\nu}$, 
even if we relax our assumption on the parity conservation. 
Furthermore by expressing our results in spinor helicity formalism we can show that if,
for two representations of the same helicity, the 3-point amplitudes for a given total number of derivatives do exist on both sides, then we do expect these particular cubic interactions to be dual to each other,  as the 3-point amplitudes in spinor helicity formalism only depend on the helicity.

To obtain, within this framework, further information about the underlying field theories, e.g. the Jacobi identity satisfied by the coupling constants in a
Yang-Mills-like theory, it is necessary to extend our  method and investigation
 to four- or higher-point amplitudes. {Also, it is interesting to relax the assumption of parity conservation to study  the 3-point amplitudes for both 
 totally  symmetric fields and the mixed-symmetric fields.}
We will report our findings  in a forthcoming paper.

\section{Appendix}
\label{appendix}
In this Appendix, we shall use on-shell gauge invariance to determine the possible 3-point amplitudes of totally symmetric polarizations. Consider three massless particles whose polarization tensors
$\epsilon_1(p_1)$, $\epsilon_2(p_2)$, $\epsilon_3(p_3)$ $(p_1+p_2+p_3=0)$
are given by,
\begin{equation}
  \epsilon^{\pm}_1(p_1)
    =\bigotimes_{i=1}^{s_1}e^{\pm}_1(p_1),\quad \epsilon^{\pm}_2(p_2)
    =\bigotimes_{i=1}^{s_2}e^{\pm}_2(p_2),\quad \epsilon^{\pm}_3(p_3)
    =\bigotimes_{i=1}^{s_3}e^{\pm}_3(p_3)
\end{equation}
where $s_i$ are the spins of the particles and {$e_i^\pm(p_i)$}  the polarization vectors.

A general Lorentz invariant  amplitude of these three particles
is a homogeneous function of $e_1$, $e_2$, $e_3$ and the momenta of degree $s_1$, $s_2$, $s_3$ and $N$, respectively, which has the form
\begin{equation}\label{amplitude}
    A(\epsilon_1,\epsilon_2,\epsilon_3;N)=\sum_{(l,m,n)\in E}\lambda_{l,m,n}(e_2\cdot e_3)^l(e_1\cdot e_3)^m (e_1\cdot e_2)^n(e_1\cdot p_2)^{s_1-m-n}(e_2\cdot p_1)^{s_2-l-n}(e_3\cdot p_1)^{s_3-l-m}
\end{equation}
{where $N$ satisfies $2|(s_1+s_2+s_3+N)$(that is, $(s_1+s_2+s_3+N)$ is even), $N\leqslant s_1+s_2+s_3$,}  and
$$
E =\left\{(l,m,n)\,\vrule\, m,n,l\in \mathbb{N},l+m+n
  =\frac{s_1+s_2+s_3-N}{2},m+n\leqslant s_1,l+n\leqslant s_2,l+m\leqslant s_3
   \right\}
$$
We have included in the amplitude only  $N$~number of derivatives 
 because the gauge invariance of the linear combination
$\sum_N \kappa_N A(\epsilon_1,\epsilon_2,\epsilon_3; N)$
is equivalent to the invariance of each
$A(\epsilon_1,\epsilon_2,\epsilon_3; N)$.
We claim that the necessary and sufficient condition for the existence of a
non-vanishing gauge invariant amplitude
$A(\epsilon_1,\epsilon_2,\epsilon_3;N)$ is
$$
   N\geqslant s_1+s_2+s_3-2\min(s_1,s_2,s_3)
$$
The proof is as follows.

To simplify the notation, we define
\begin{equation}
     \overline{l,m,n}\equiv(e_2\cdot e_3)^l(e_1\cdot e_3)^m (e_1\cdot e_2)^n
     (e_1\cdot p_2)^{s_1-m-n}(e_2\cdot p_1)^{s_2-l-n}(e_3\cdot p_1)^{s_3-l-m}
\end{equation}
then the amplitude~(\ref{amplitude}) becomes
\begin{equation}\label{simplifiedAmplitude}
    A(\epsilon_1,\epsilon_2,\epsilon_3;N)
    =\sum_{(l,m,n)\in E}\lambda_{l,m,n}\cdot\overline{l,m,n}
\end{equation}
The constraints on the integers $l,m,n$ in the set $E$ are equivalent to the following conditions:
\begin{align}
      n & =\frac{s_1+s_2+s_3-N}{2}-l-m\equiv n(l,m)\label{condition1}\\
     l & \geqslant \max\left(0,\frac{s_1+s_2+s_3-N}{2}-s_1\right)
          \label{condition2}\\
     m & \geqslant \max\left(0,\frac{s_1+s_2+s_3-N}{2}-s_2\right)
          \label{condition3}\\
     l & +m\leqslant \min\left(s_3,\frac{s_1+s_2+s_3-N}{2}\right)~.
          \label{condition4}
\end{align}
We can rewrite the above inequalities in the form
\begin{equation}\label{rangeLM}
 \left\{
    \begin{aligned}
    l_{\min}^N\equiv & \max\left(0,\frac{s_1+s_2+s_3-N}{2}-s_1\right)\leqslant l
     \leqslant \min\left(s_3,\frac{s_1+s_2+s_3-N}{2}\right)\equiv l_{\max}^N    \\
      m_{\min}^N\equiv & \max\left(0,\frac{s_1+s_2+s_3-N}{2}-s_2\right)\leqslant
      m\leqslant\min\left(s_3,\frac{s_1+s_2+s_3-N}{2}\right)-l\equiv m_{\max}^N(l)
    \end{aligned}
 \right.
\end{equation}
Now the equation~(\ref{simplifiedAmplitude}) becomes
\begin{equation}
    A(\epsilon_1,\epsilon_2,\epsilon_3;N)=\sum_{l=l_{\min}^N}^{l_{\max}^N}\sum_{m=m_{\min}^N}^{m_{\max}^N(l)}\lambda_{l,m,n}\cdot\overline{l,m,n}
        =\sum_{l=l_{\min}^N}^{l_{\max}^N}A_l
\end{equation}
where
$$
  A_l\equiv\sum_{m=m_{\min}^N}^{m_{\max}^N(l)}\lambda_{l,m,n}\cdot\overline{l,m,n}
$$
The amplitude is invariant under the gauge transformation,
$\delta_1 e_1= p_1,\delta_1 e_2=\delta_1 e_3=0$,
(the notation $\delta_i e_j$ being defined in~(\ref{gaugeT})).
Because $\delta_1 A_l$ has the structure
$$
\delta_1 A_l=\sum_m\lambda'_{l,m,n}\cdot\overline{l,\times,\times}
$$
where $\times$ stands for some numeric constants, 
we have
$$
  \delta_1 A=\sum_l \delta_1 A_l=0\,\,\Leftrightarrow\,\,
     \delta_1 A_l=0\quad(\text{for all }l)
$$
Therefore
\begin{align}
    \delta_1 A_l
      &=\sum_{m=m_{\min}^N}^{m_{\max}^N(l)}(m\lambda_{l,m,n}
        \cdot\overline{l,m-1,n}+n\lambda_{l,m,n}\cdot\overline{l,m,n-1})\\
    &= \sum_{m=m_{\min}^N-1}^{m_{\max}^N(l)-1}(m+1)\lambda_{l,m+1,n-1}
      \cdot\overline{l,m,n-1}
      +\sum_{m=m_{\min}^N}^{m_{\max}^N(l)}n\lambda_{l,m,n}
      \cdot\overline{l,m,n-1}\\
    & =\sum_{m=m_{\min}^N}^{m_{\max}^N(l)-1}\big[(m+1)
    \lambda_{l,m+1,n-1}+n\lambda_{l,m,n}\big]\cdot\overline{l,m,n-1}\\
    & \quad+m_{\min}^N\lambda_{l,m_{\min}^N,n(l,m_{\min}^N)}
      \cdot\overline{l,m_{\min}^N-1,n(l,m^N_{\min})}\\
    & \quad+n(l,m_{\max}^N(l))\lambda_{l,m_{\max}^N(l),n(l,m_{\max}^N(l))}
      \cdot\overline{l,m_{\max}^N(l),n(l,m_{\max}^N(l))-1}\\
    & =0
\end{align}
which leads to
\begin{equation}
\label{proportionality}
   \left\{\begin{aligned}
  0 & =(m+1)\lambda_{l,m+1,n-1}+n\lambda_{l,m,n}\quad(m_{\min}^N\leqslant
       m\leqslant m_{\max}^N(l)-1)\\
  0 & =m_{\min}^N\lambda_{l,m_{\min}^N,n(l,m_{\min}^N)}\\
  0 & =n(l,m_{\max}^N(l))\lambda_{l,m_{\max}^N(l),n(l,m_{\max}^N(l))}
   \end{aligned}\right.
\end{equation}
Now choose $l$ so that $A_l\neq 0$, then we must require
\begin{equation}
 \label{boundary}
    m_{\min}^N=n(l,m_{max}^N(l))=0~,
\end{equation}
otherwise we would have, $\lambda_{l,m_{\min}^N,n(l,m_{\min}^N)}=0$, or,
$\lambda_{l,m_{\max}^N(l),n(l,m_{\max}^N(l))}=0$,
and it follows from the recursion relation,
$(m+1)\lambda_{l,m+1,n-1}+n\lambda_{l,m,n}=0$,
that all $\lambda_{l,\times,\times}$ vanish.
Equations~(\ref{condition1}), (\ref{rangeLM}) and (\ref{boundary}) yield
\begin{align}
    & \frac{s_1+s_2+s_3-N}{2}-s_2\leqslant 0\label{inequality2}\\
    & \frac{s_1+s_2+s_3-N}{2}-s_3\leqslant 0\label{inequality3}
\end{align}
Similarly, by performing a gauge transformation on particle 2 (or 3) we obtain
\begin{equation}\label{inequality1}
    \frac{s_1+s_2+s_3-N}{2}-s_1\leqslant 0
\end{equation}
The inequalities~(\ref{inequality2}),~(\ref{inequality3}), and (\ref{inequality1}) can be combined into
\begin{equation}\label{range_of_N}
    N\geqslant s_1+s_2+s_3-2\min(s_1,s_2,s_3)
\end{equation}
which is what we intend to prove.
Next we will show that for each $N$ satisfying the above inequality, the amplitude $A(s_1,s_2,s_3;N)$ is unique (up to an overall factor).

With the help of $(\ref{range_of_N})$, the set $E$ becomes
\begin{equation}
E=\left\{(l,m,n)\,\vrule\, m,n,l\in \mathbb{N},l+m+n
  =\frac{s_1+s_2+s_3-N}{2}\right\}
\end{equation}
According to equation (\ref{proportionality}) we have
\begin{equation}
    \lambda_{l,m,n}\propto \lambda_{l,m+1,n-1}
\end{equation}
Similarly, by performing a gauge transformation on particle 2 we have
\begin{align}
    & \lambda_{l,m,n}\propto \lambda_{l+1,m,n-1}
\end{align}
From which we obtain
$$
    \lambda_{l,m,n}\propto \lambda_{l,0,M-l}\propto \lambda_{l',0,M-l'}
      \propto \lambda_{l',m',n'}
$$
where $\lambda_{l,m,n}$ and $\lambda_{l',m',n'}$ are any two coefficients and
$$
    M=\frac{s_1+s_2+s_3-N}{2}~.
$$
Therefore all the coefficients are proportional to each other.
As a result, the amplitude, $A(\epsilon_1,\epsilon_2,\epsilon_3;N)$,
is unique up to an overall factor.

It turns out that there are four basic gauge invariant amplitudes, namely, the amplitudes $e_1\cdot p_2$, $e_2\cdot p_1$, $e_3\cdot p_1$ of a spin-1 particle (whose polarization vector is $e_i,i=1,2,3$) scattering with two scalar particles and the Yang-Mills amplitude
$A_{\text{YM}}$:
\begin{equation}
        A_{\text{YM}}(e_1,e_2,e_3)\equiv (e_1\cdot p_2)(e_2\cdot e_3)-(e_2\cdot p_1)(e_1\cdot e_3)+(e_3\cdot p_1)(e_1\cdot e_2)~,
\end{equation}
from which a general gauge invariant amplitude of particles whose polarizations
are $\epsilon_1,\epsilon_2$ and $\epsilon_3$ can be uniquely constructed:
\begin{equation}
    \begin{aligned}
        & A(\epsilon_1,\epsilon_2,\epsilon_3;N)\\
        = & (e_1\cdot p_2)^{\frac{s_1-s_2-s_3+N}{2}}(e_2\cdot p_1)^{\frac{s_2-s_1-s_3+N}{2}}(e_3\cdot p_1)^{\frac{s_3-s_1-s_2+N}{2}}A_{\text{YM}}(e_1,e_2,e_3)^{\frac{s_1+s_2+s_3-N}{2}}
    \end{aligned}
\end{equation}
where the exponents are determined by the requirement that the amplitude have the correct numbers of polarization vectors and momenta:
$\#(e_1)=s_1,\#(e_2)=s_2,\#(e_3)=s_3,\sum_i\#(p_i)=N$.

\section*{Acknowledgements}
We would like to thank  George Savvidy for his valuable comments on an earlier draft.
Many thanks to {Chen Gang,} {Zepeng He,} {Heyang Long,} {Tianheng Wang} 
and {Xincheng Yu} for many useful discussions and the great fun. \\
This research project has been supported in parts by the NSF China under Contract 
No.~11775110,  No.~11690034  and No.~11405084.
We also acknowledge the European Union's Horizon 2020 Research and Innovation~(RISE) programme under the Marie Sk\'lodowska-Curie grant agreement No.~644121,
and  the Priority Academic Program Development for
Jiangsu Higher Education Institutions (PAPD).

\providecommand{\href}[2]{#2}
\begingroup
\raggedright

\endgroup

\end{document}